\documentclass[journal, hidelinks]{IEEEtran}
\usepackage[nocompress]{cite}
\usepackage{graphicx}
\usepackage[T1]{fontenc}
\usepackage[tt=false, type1=true]{libertine}
\usepackage[varqu]{zi4}
\usepackage[libertine]{newtxmath}

\usepackage{amsmath}
\usepackage{comment}
\usepackage{tabularx}
\usepackage{multirow, multicol}
\newcolumntype{Y}{>{\centering\arraybackslash}X} 
\usepackage{tabularray}
\usepackage{makecell}
\usepackage{array}
\usepackage{xspace}
\usepackage[font=small]{caption}
\usepackage{orcidlink}
\usepackage{nicefrac}
\usepackage{siunitx}
\usepackage{float}
\usepackage{dblfloatfix}
\usepackage{enumitem}
\usepackage{subcaption}
\usepackage{relsize}
\usepackage{adjustbox}
\usepackage{soul}
\usepackage{amsthm}
\usepackage{mathtools}
\usepackage{bbm}
\usepackage{etoolbox}
\usepackage{titlesec}
\usepackage{diagbox}
\usepackage[hang, flushmargin]{footmisc}
\usepackage{algorithm}
\usepackage[noend]{algpseudocode}

\makeatother

\def\sothat{\,\vert\,}
\newcommand\abs[1]{\left\lvert#1\right\rvert}

\newcommand{\kk}{\text{\scriptsize\sf k}\xspace}
\newcommand{\MM}{\text{\scriptsize\sf M}\xspace}

\DeclareMathOperator*{\argmax}{argmax}

\newcommand{\elevate}[2][1mm]{\raisebox{#1}{\({#2}\)}}

\def\hedge{\mbox{h\!\!\;-\!\!\;edge}\xspace}
\def\hgraph{\mbox{h\!\!\;-\!\!\;graph}\xspace}
\def\hedges{\mbox{h\!\!\;-\!\!\;edges}\xspace}
\def\hgraphs{\mbox{h\!\!\;-\!\!\;graphs}\xspace}

\def\Hgraph{\mbox{H\!\!\;-\!\!\;graph}\xspace}
\def\Hedges{\mbox{H\!\!\;-\!\!\;edges}\xspace}

\algnewcommand\ProcedureLine[1]{\item[\textbf{Procedure:}] #1}

\newskip\paragraphbreak
\titleformat{\subsubsection}[runin]
  {\itshape} 
  {\thesubsubsection.}{0.5em}{}[: \quad] 

\titlespacing*\section{0pt}{*2}{*1}
\titlespacing*\subsection{0pt}{*2}{*1}
\titlespacing{\subsubsection}{0pt}{4pt}{0pt} 

\expandafter\def\expandafter\normalsize\expandafter{%
    \normalsize
    \setlength\abovedisplayskip{5pt} 
    \setlength\belowdisplayskip{5pt} 
    \setlength\abovedisplayshortskip{3pt} 
    \setlength\belowdisplayshortskip{3pt} 
}

\newlength{\itemizepadding}
\setlist[itemize]{topsep=\itemizepadding, leftmargin=3em} 
\setlist[enumerate]{topsep=\itemizepadding, leftmargin=3em} 

\IEEEoverridecommandlockouts
\IEEEaftertitletext{\vspace{-11pt}}

\renewcommand{\IEEEbibitemsep}{0pt plus 0.5pt}
\makeatletter
\renewcommand{\@biblabel}[1]{[#1]\hfill} 
\IEEEtriggercmd{\reset@font\normalfont\fontsize{8pt}{9pt}\selectfont} 
\IEEEtriggeratref{1}
\makeatother

\usepackage{nicematrix, tikz, calc, relsize}
\usetikzlibrary{calc}

\newcommand{\slantedssnstableall}{
\begin{table}[t]
    \vspace{13pt}
    \centering
    \renewcommand{\arraystretch}{1.1} 
    \resizebox{\columnwidth}{!}{%
        \begin{minipage}{1.25\columnwidth}
            \centering
            \setlength{\tabcolsep}{0pt}
            \begin{NiceTabularX}{\textwidth}{X[1.4,c,m] *{12}{X[0.9,c,m]}}
                & & & & & & & & & & & & \\
                $\abs{N}$
                & 20\kk & 110\kk & 216\kk & 302\kk & 14\kk & 208\kk & 194\kk & 6.9\MM & 231\kk & 16\kk & 64\kk & $256\kk$ \\
                $\sum_{e \in E} \abs{e}$
                & 766\kk & 23\MM & 90\MM & 256\MM & 875\kk & 145\MM & 133\MM & 577\MM & 70\MM & 2.1\MM & 12.6\MM & 67.4\MM \\
                $\text{avg}_{e \in E} \abs{e}$
                & 37.3 & 210.3 & 417.2 & 848.1 & 63.2 & 696.2 & 688.3 & 83.5 & 304.7 & 128 & 192 & 256 \\
                \textbf{$\Omega, \, \Delta$}
                & ${\scriptstyle 2}^{10}\!\!, {\scriptstyle 2}^{12}$ & ${\scriptstyle 2}^{10}\!\!, {\scriptstyle 2}^{12}$
                & ${\scriptstyle 2}^{12}\!\!, {\scriptstyle 2}^{16}$ & ${\scriptstyle 2}^{12}\!\!, {\scriptstyle 2}^{16}$
                & ${\scriptstyle 2}^{10}\!\!, {\scriptstyle 2}^{12}$ & ${\scriptstyle 2}^{12}\!\!, {\scriptstyle 2}^{16}$
                & ${\scriptstyle 2}^{12}\!\!, {\scriptstyle 2}^{16}$ & ${\scriptstyle 2}^{12}\!\!, {\scriptstyle 2}^{16}$
                & ${\scriptstyle 2}^{12}\!\!, {\scriptstyle 2}^{16}$ & ${\scriptstyle 2}^{10}\!\!, {\scriptstyle 2}^{12}$
                & ${\scriptstyle 2}^{10}\!\!, {\scriptstyle 2}^{12}$ & ${\scriptstyle 2}^{10}\!\!, {\scriptstyle 2}^{12}$ \\
                \CodeAfter
                \begin{tikzpicture}
                    \foreach \j in {1,2,...,14}
                        \draw (2-|\j) -- (6-|\j);
                    \foreach \i in {2,...,13}
                        \draw (2-|\i) -- ++(155:1.5cm); 
                    \foreach \label [count=\i from 1] in {
                        \textsmaller[1]{16\kk-model}, \textsmaller[1]{64\kk-model}, \textsmaller[1]{256\kk-model}, \textsmaller[1]{1\MM-model},
                        \textsmaller[1]{lenet}, \textsmaller[1]{alexnet}, \textsmaller[1]{\smash[b]{vgg11}}, \textsmaller[1]{mobilenet},
                        \textsmaller[1]{allen\! v1}, \textsmaller[1]{16\kk-rand}, \textsmaller[1]{64\kk-rand}, \textsmaller[1]{256\kk-rand}
                    } {                            
                        \node[rotate=-25, anchor=south east, inner sep=-1pt] 
                            at ($ (2-|\inteval{\i+1}) + (3pt,3pt) $) {\label};
                    }
                    \draw (2-|14) -- ++(155:1.5cm);
                    \draw (2-|2) -- (2-|14);
                \end{tikzpicture}
            \end{NiceTabularX}
        \end{minipage}%
    }
    \vspace{-2pt}
    \caption{Spiking neural networks used in the experiments \cite{BenchmarkSNNs}.}
    \vspace{-4pt}
    \label{tab:snns_all}
\end{table}
}

\newcommand{\slantedispdtable}{
\begin{table*}[t]
    \vspace{13pt}
    \centering
    \renewcommand{\arraystretch}{1.1} 
    \resizebox{\textwidth}{!}{%
        \begin{minipage}{1.25\textwidth} 
            \centering
            \setlength{\tabcolsep}{0pt}
            \begin{NiceTabularX}{\textwidth}{X[1.4,c,m] *{18}{X[0.9,c,m]}}
                & & & & & & & & & & & & & & & & & & \\
                $\abs{N}$
                & 204\kk & 314\kk & 370\kk & 440\kk & 470\kk & 520\kk & 735\kk & 821\kk
                & 854\kk & 1.11\MM & 1.13\MM & 1.14\MM & 1.35\MM & 2.36\MM & 2.59\MM & 2.94\MM & 2.97\MM & 3.37\MM \\                
                $\sum_{e \in E} \abs{e}$
                & 808\kk & 1.30\MM & 1.49\MM & 1.69\MM & 2.03\MM & 2.05\MM & 2.81\MM & 3.29\MM
                & 3.56\MM & 4.76\MM & 4.49\MM & 5.09\MM & 5.72\MM & 8.75\MM & 11.45\MM & 12.46\MM & 13.76\MM & 13.12\MM \\
                $\text{avg}_{e \in E} \abs{e}$
                & 3.58 & 4.15 & 3.41 & 3.31 & 4.46 & 3.68 & 3.65 & 4.07
                & 3.65 & 3.96 & 3.45 & 4.12 & 3.58 & 3.58 & 3.84 & 4.10 & 4.54 & 4.06 \\
                \CodeAfter
                \begin{tikzpicture}
                    \foreach \j in {1,2,...,20}
                        \draw (2-|\j) -- (5-|\j);
                    \foreach \i in {2,...,19}
                        \draw (2-|\i) -- ++(155:0.7cm); 
                    \foreach \label [count=\i from 1] in {
                        \textsmaller[1]{01}, \textsmaller[1]{02}, \textsmaller[1]{03}, \textsmaller[1]{04},
                        \textsmaller[1]{05}, \textsmaller[1]{06}, \textsmaller[1]{07}, \textsmaller[1]{08},
                        \textsmaller[1]{09}, \textsmaller[1]{10}, \textsmaller[1]{11}, \textsmaller[1]{12},
                        \textsmaller[1]{13}, \textsmaller[1]{14}, \textsmaller[1]{15}, \textsmaller[1]{16},
                        \textsmaller[1]{17}, \textsmaller[1]{18}
                    } {                            
                        \node[rotate=-25, anchor=south east, inner sep=-1pt] 
                            at ($ (2-|\inteval{\i+1}) + (9pt,3pt) $) {\label};
                    }
                    \draw (2-|20) -- ++(155:0.7cm);
                    \draw (2-|2) -- (2-|20);
                \end{tikzpicture}
            \end{NiceTabularX}
        \end{minipage}%
    }
    \vspace{-2pt}
    \caption{ISPD98 hypergraphs scaled by $16\times$ used in the experiments \cite{ISPD98}.}
    \vspace{-14pt} 
    \label{tab:ispd}
\end{table*}
}

\usepackage{booktabs}

\newcommand{\complexitytable}{
\begin{table}[t]
    \vspace{5pt}
    \centering
    \renewcommand{\arraystretch}{1.1} 
    \resizebox{\columnwidth}{!}{%
        \begin{minipage}{1.25\columnwidth}
            \centering
            \setlength{\tabcolsep}{0pt}
            \begin{NiceTabularX}{0.8\textwidth}{X[2.8,l,m] *{2}{X[1.2,c,m]}}
                \toprule
                \textbf{Algorithm Step} & \textbf{Work} & \textbf{Span} \\
                \midrule
                Candidate Pairs Proposal & $\abs{N} \cdot h \cdot d$ & $h$ \\
                Nodes Matching & $\abs{N}$ & $1$ \\
                Coarse \Hgraph Construction & $\abs{N} \cdot h \cdot d + \abs{E} \cdot d$ & $1$ \\
                Refinement Gain Calculation & $\abs{N} \cdot h \cdot d + \abs{N} \cdot \abs{P}$ & $h$ \\
                Moves Sequence Construction & $\abs{N} \cdot \log\,\abs{N}$ & $\log\,\abs{N}$ \\
                Events Validity Check & $\abs{N} \cdot \log\,\abs{N} \cdot h$ & $\log\,\abs{N}$ \\
                First Neighbors Construction & $\abs{N} \cdot h \cdot d$ & $h$ \\
                \bottomrule
            \end{NiceTabularX}
        \end{minipage}%
    }
    \vspace{-1pt} 
    \caption{Summary of work and span of every algorithm step.}
    \vspace{-4pt}
    \label{tab:complexity}
\end{table}
}

\begin{document}

\title{Hypergraph Partitioning on GPU with\\Distinct Incident Hyperedges and Size Constraints}

\author{
    Marco Ronzani \orcidlink{0009-0002-8485-0717}\,, Cristina Silvano \orcidlink{0000-0003-1668-0883}\,,\;\;DEIB, Politecnico di Milano, Italy\vspace{-18pt}
}

\markboth{Submitted to IEEE Transactions on Parallel and Distributed Systems}%
{Hypergraph Partitioning on GPU with Distinct Incident Hyperedges and Size Constraints}

\maketitle

\begin{abstract}
    Hypergraph partitioning is a recurring NP-hard problem in engineering; its efficient solution at scale hinges on parallelism.
    This work proposes a GPU-centric algorithm for multi-level hypergraph partitioning aimed at a specific set of problem constraints: limited size and distinct inbound hyperedges per partition.
    Manipulating hypergraphs requires deeply nested traversals and concurrent decision-making; our constraints impose further set operations amidst that.
    In turn, we design algorithms around the GPU's hierarchical parallelism and our problem's specifics.
    When forming partitions, we materialize the hypergraph's incidence structure and unique neighborhoods in memory to exploit set sparsity and batch node-pairing scores in shared memory.
    Upon refining partitions, we chain node moves into improving paths and cycles, checking their validity via cumulative set size variations reduced in parallel over moves.
    Thus, our dominant kernels exhibit a span linear in local hypergraph parameters.
    Results show an average $\mathbf{380\times}$ speedup and a $\mathbf{1.2}$-$\mathbf{2.0\times}$ reduction in connectivity compared to a sequential multi-level partitioner.
    With minor changes, we also support k-way balanced partitioning, running $\mathbf{5\times}$ faster than CPU methods with a $\mathbf{\sim\!5\%}$ quality loss for $\mathbf{k\!=\!2}$, outperforming an existing GPU partitioner at comparable runtime, \mbox{with no measurable overhead} from the added constraints handling logic.%
\end{abstract}

\vspace{-3pt}

\begin{IEEEkeywords}
Hypergraph partitioning, parallel algorithms, GPU acceleration, incidence constraint, size constraint.
\end{IEEEkeywords}

\vspace{-3pt} 

\section{Introduction}

Hypergraph partitioning is a widespread problem throughout computer science, from VLSI to scientific and high-performance computing.
Being NP-hard, practical solvers rely on heuristics that increasingly trade quality of results for time as instance size grows \cite{AdvancesInHypergraphPartitioning, PartitioningHypergraphsIsHard}.
For this reason, the efficient, massively parallel implementation of hypergraph partitioning algorithms holds the potential for time savings and performance improvements across many domains.
However, due to the sparse and irregular structure of hypergraphs, such algorithms are far from trivially parallelizable \cite{AdvancesInHypergraphPartitioning, gHyPart}.

In this work, we develop a \textbf{GPU-parallel algorithm} for \textbf{directed hypergraph partitioning} under size and incidence constraints.
Each partition is limited in the number of nodes it can contain and in the number of its distinct inbound hyperedges.
The goal of partitioning is to minimize the connectivity, the total weight of cuts induced by hyperedges between partitions.

This particular formulation emerges in several settings.
In mapping Spiking Neural Networks (SNNs) to neuromorphic hardware, constraints reflect the limited resources of hardware cores, while a lower connectivity reduces spike traffic and transmission costs \cite{AxonFlow, MappingVeryLargeSNNtoNHW}.
In VLSI and FPGAs, finite I/O often bounds incident connections while less communication saves time and energy \cite{hMETIS_vlsi, AdvancesInHypergraphPartitioning}.
A notable mention is chiplets-based designs, that require partitioning modules across multiple dies with a tight interface budget \cite{ChipletsCombinatorics}.
Other applications are workload distribution in supercomputers under limited interconnect bandwidth \cite{ChallengesInDynamicLoadBalancing} and the optimization of parallel algorithms, like the sparse matrix-multiply kernel \cite{HypergraphForSparseMatMul, gHyPart}.
In many of these cases, the scale of hypergraphs involved is rapidly growing past millions of nodes and hyperedges, totaling billions of pins.

By their own nature, working with hypergraphs involves nested iterations.
Nodes are incident to multiple hyperedges, each containing dozens of pins.
Thus, already with neighborhood traversals -- going from a node to its incident hyperedges, then to their pins -- hop count quickly explodes.
And several such visits form the basis for partitioning \cite{AcceleratedCoarseningProcedure, SIMDefficientGraphsOnGPU}.
Our problem settings further amplify this cost with the addition of loop nests that check bounds on the number of hyperedges entering a partition.
Tracking which requires repeated set unions, intersections, and deduplication.

Although traversals themselves are trivially parallel, hypergraph partitioning algorithms do not align well with parallel hardware.
For one, involved heuristics often rely on sequential decision making and backtracking, that result in highly contended atomic operations and heavy synchronization \cite{HyperG}.
Even then, a hypergraph's incidence structure is typically irregular and hyperedges are unevenly distributed.
Hence, achieving optimal workload distribution would presuppose that the hypergraph is already well partitioned \cite{AdvancesInHypergraphPartitioning}.

To this day, graph and hypergraph partitioning on CPU has been extensively studied \cite{AdvancesInHypergraphPartitioning}, notable sequential works being hMETIS \cite{hMETIS_k_way}, KaHyPar \cite{KaHyPar}, and PaToH \cite{PaToH}, with also multi-threaded developments such as Mt-KaHyPar \cite{MtKaHyPar}, BiPart \cite{BiPart}, and Zoltan \cite{Zoltan}.
Moving to GPU, efficient hypergraph-wide optimal updates have been devised to compete with complex sequential heuristics.
First G-kway \cite{Gkway}, then HyperG \cite{HyperG}, implemented arbitrary ordering techniques to enable both parallel coarsening and refinement.
Instead, gHyPart \cite{gHyPart} explored adaptive parallelization strategies to accommodate changes in the structure of input hypergraphs.
Yet, none of these approaches consider incidence constraints, nor have been tested at realistic scales beyond 10\MM pins \cite{ChipletsCombinatorics, MappingVeryLargeSNNtoNHW}.

The algorithm we implement is based on the multi-level approach \cite{hMETIS_k_way, FMpartitioning}.
In it, nodes that participate in similar sets of hyperedges are progressively clustered during a coarsening phase, until further aggregation would violate constraints.
The resulting clusters define an initial partitioning, which is then refined by uncoarsening the hypergraph and evaluating node moves between partitions.

All throughout the above process, the hypergraph and its partitions must stay valid; therefore, two steps in particular are stressed by our set of constraints.
Coarsening must only form valid clusters, meaning every pair of neighbors needs the intersection of their inbound hyperedges set to be computed.
Refinement must land on a valid state for every partition, while still allowing violations in between improving node moves.
Both such constraint checks involve tracking inbound sets inside of already deeply nested operations.

To accommodate this, we redesign the multi-level scheme around the GPU's hierarchical execution model.
Every traversal and set operation is handled cooperatively by warps and threads.
Data structures exploit the hypergraph's sparsity while preserving data access locality.
Finally, every optimization problem is rethought such that nodes can decide independently and agree on valid actions later.

\subsection{Contributions}

In this work, we detail a novel GPU-based multi-level deterministic hypergraph partitioner that handles constraints on partition size and distinct inbound hyperedges per partition.
To the best of our knowledge, this is the first massively parallel implementation of algorithms specifically addressing such constraints.
In particular, our algorithms feature:
\begin{enumerate}
    \item a constraint-aware coarsening strategy that estimates inbound set union size inline during neighbor scoring;
    \item an exact parallel dynamic programming algorithm for maximum-weight matching on the pseudo-forest induced by proposed coarsening pairs;
    \item a refinement strategy that enables the simultaneous application of interfering node moves by organizing them into gain-ranked feasible paths and swap cycles;
\end{enumerate}
Furthermore, to implement these ideas efficiently on GPU, we introduce:
\begin{enumerate}
    \item a hierarchical mapping of nested hypergraph traversals onto blocks, warps, and threads, confining sequential work to local structural parameters and yielding an effective span linear in node degree;
    \item the precomputation of deduplicated neighborhoods, enabling batched neighbor scoring in shared memory;
    \item a sparse event-based method to validate size and inbound hyperedge constraints in parallel for all refinement node moves, using fast prefix sums and sorting primitives;
\end{enumerate}
The resulting algorithm, tested on SNNs with up to 500\MM pins, is on average $380\times$ faster than a sequential multi-level equivalent, while yielding a $1.2$–$2.0\times$ reduction in cut cost, an improvement of up to $2.4\times$ over existing SNN partitioners.

With minimal changes, our approach is also applicable to the $k$-way balanced partitioning problem.
We thus evaluate it on augmented versions of the ISPD98 benchmark hypergraphs \cite{ISPD98}, achieving a mean $5\times$ speedup over a SoTA multi-threaded CPU partitioner with a $5\%$ cut-net increase on $k\!=\!2$.
This yields better quality than a SoTA GPU partitioner at comparable runtime, despite the added constraints logic.

All stated contributions represent major improvements over the prototype of our partitioner presented in \cite{AxonCUDA-IPDPS}.
Namely, in-histogram inbound set size tracking, exact matching, and moves chaining into paths and cycles.
As part of our results, we demonstrate the impact of all such algorithmic design choices and parameter settings over coarsening quality, refinement effectiveness, and~runtime~efficiency.

\vspace{6pt}

\noindent
The remainder of this article is structured as follows.
Sec.~\ref{sec:problem_definition} formalizes the partitioning problem, while Sec.~\ref{sec:multi_level_scheme} introduces our variant of the multi-level scheme.
Sec.~\ref{sec:hypergraph_handling} presents the design principles underlying our implementation.
Secs.~\ref{sec:coarsening} and \ref{sec:refinement} then detail our parallel algorithms.
Our experiments are reported throughout Sec.~\ref{sec:experimental_results}.
Finally, Sec.~\ref{sec:conclusion} gives a few conclusive remarks.

\section{Problem Definition}\label{sec:problem_definition}



\subsection{Hypergraph Partitioning Model}\label{subsec:hgraph_part_model}

Hypergraphs (\hgraphs) are a generalization of graphs where edges can connect more than two nodes, thereby becoming hyperedges (\hedges).
For us, a weighted directed hypergraph $G(N, E, \omega)$ consists of a set $N$ of nodes and a set $E$ of hyperedges.
Each \hedge $e \in E$ contains a subset of nodes (or pins) $e \subseteq N$, with $src(e)$ isolating the \hedge's sources and $dst(e)$ its destinations.
Owing to $e$ being a set, we assume no duplicate pins nor self-cycles, $\forall e, \: src(e) \cap dst(e) = \varnothing$.
The function $\omega : E \rightarrow \mathbb{R}$ assigns a weight to each \hedge.
In addition, we define incidence sets $in(n) = \{e \in E \sothat n \in dst(e)\}$ and $out(n) = \{e \in E \sothat n \in src(e)\}$ as the sets of inbound and outbound \hedges from a node $n \in N$, and $\mathcal{I}(n) = in(n) \cup out(n)$ as the node's set of incident \hedges.
Lastly, we denote the neighbors of a node $n \in N$ as $\mathcal{N}(n) = \{m \in e \sothat e \in \mathcal{I}(n)\} \setminus \{n\}$.
That being the subset of nodes partaking in any \hedge together with $n$.

A partitioning of $G$ is a set $P \subset \mathcal{P}(N)$ of pairwise disjoint subsets -- partitions -- of its nodes such that $\bigcup_{p \in P} p = N$.
With $\mathcal{P}(\cdot)$ denoting the power set.
Equivalently, a partitioning can be represented by a function $\rho : N \rightarrow P$ assigning each node to a partition, where $\rho(n) = p$ if and only if $n \in p$.

For convenience, we assume an arbitrary total order to exist over nodes $\prec_{id}N$, \hedges $\prec_{id}E$, and partitions $\prec_{id}P$, induced by their id-based representation in memory.

\subsection{Objective and Constraints}\label{subsec:objective_and_constraints}

Our partitioning constraints pose hard limits on the number of nodes and of distinct inbound \hedges per partition.
Let $\Omega$ be the maximum size of a partition, from which $\forall p \in P, \: \abs{p} \leq \Omega$.
Let $\Delta$ be the maximum number of allowed distinct inbound \hedges to a partition, hence $\abs{\bigcup_{n \in p} in(n)} \leq \Delta$.
Furthermore, let $pins(p, e) = \abs{\{n \in e \sothat \rho(n) = p\}}$ be the number of pins \hedge $e$ owns in partition $p$, and $pins_{in}(p, e) = \abs{\{n \in dst(e) \sothat \rho(n) = p\}}$ the number of times an \hedge $e$ is inbound to a partition $p$.
The second constraint can thus be written as $\forall p \in P, \: \abs{\{e \in E \sothat pins_{in}(p, e) > 0\}} \leq \Delta$.

Our objective function is the \textbf{connectivity}, the number of cuts on each \hedge times its weight.
That is, we pay once an \hedge's weight for every partition beyond the first it touches:
\begin{equation}\label{eq:connectivity}
    Conn_G(\rho) = \sum_{e \in E} \omega(e) \cdot (\abs{\{\rho(n) \sothat n \in e\}} - 1) \text{.}
\end{equation}
The goal for partitioning is to minimize connectivity subject to the above constraints.

While the remainder of this work focuses on the distinct inbound \hedges constraint, it must be noted that all presented solutions can be trivially reformulated to consider a distinct outbound or incident \hedges count constraint and undirected \hgraphs.
In addition, the present discussion assumes that a valid solution always exists.


\section{The Multi-level Partitioning Scheme}\label{sec:multi_level_scheme}


\begin{figure}[t]
    \centering
    \includegraphics[width=1.0\columnwidth]{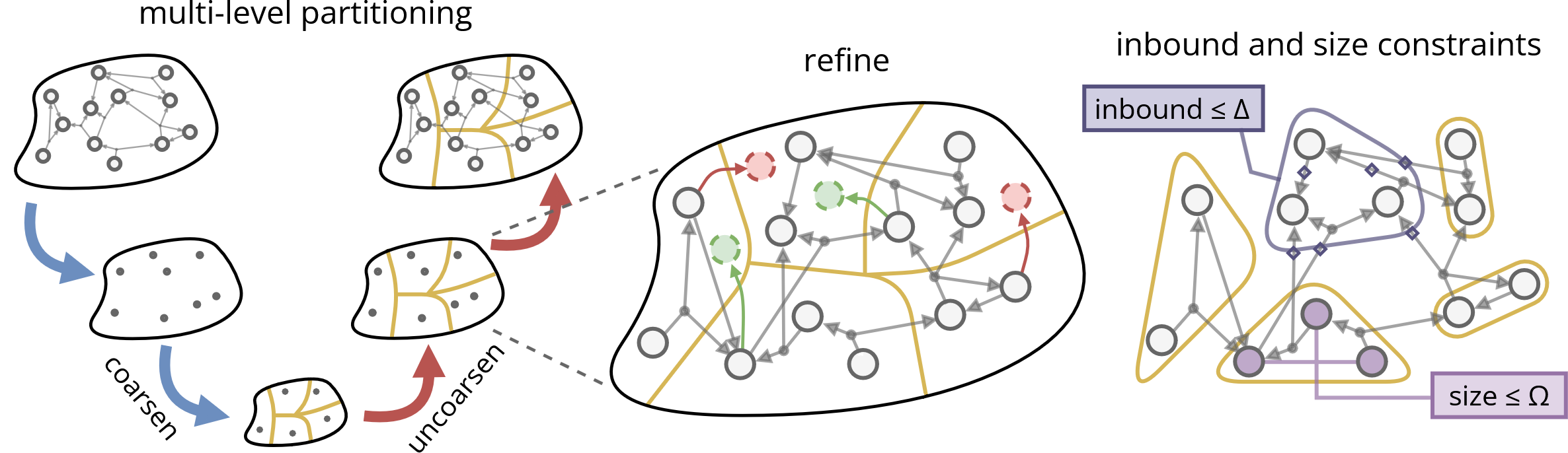}
    \vspace{-12pt}
    \caption{Overview of the multi-level hypergraph partitioning scheme.}
    \vspace{-2pt} 
    \label{fig:overview}
\end{figure}

The multi-level heuristic comprises the \textit{coarsening}, \textit{initial partitioning}, and \textit{uncoarsening} phases, as shown in Fig.~\ref{fig:overview}.
The idea is to progressively simplify the partitioning problem by coarsening nodes into larger and larger clusters, until few enough remain that a good initial partitioning can be cheaply determined.
Then, clusters are undone while localized attempts are made to improve the solution \cite{hMETIS_k_way, hMETIS_vlsi, AdvancesInHypergraphPartitioning}.
We here rethink this classic scheme in light of our constraints.
Our coarsening routine implements edge-coarsening \cite{hMETIS_vlsi}, while refinement is based on the Fiduccia–Mattheyses algorithm \cite{FMpartitioning}.
Constraint checks aside, the \hgraph is treated as undirected.

The first coarsening level takes $G(N, E, \omega)$ and constructs its coarse version $G'(N', E', \omega')$.
Where $N' \subset \mathcal{P}(N)$ is a set of coarse nodes, pairwise disjoint clusters over $N$ with $\bigcup_{n' \in N'} n' = N$.
This is complemented by a node to cluster assignment function $\gamma : N \rightarrow N'$ such that $\gamma(n) = n'$ iff $n \in n'$.
Coarse \hedges $E'$ are built accordingly as $E' = \{\{\gamma(n) \sothat n \in e\} \sothat e \in E\}$ and $\omega'(e') = \omega(e)$.
To pace the process, we limit each cluster to at most two nodes $\forall n' \in N', \: \abs{n'} \leq 2$.
As we coarsen we track the overall size of coarse nodes as $size : N \cup N' \rightarrow \mathbb{N}^+$, where $\forall n \in N, \, size(n) = 1$ and $\forall n \in N', \, size(n') = \abs{n'}$.
Clusters form the basis for partitions and must thus respect the same constraints.

The goal for coarsening is to cluster together nodes appearing in similar sets of \hedges \cite{AcceleratedCoarseningProcedure}.
That is, clusters should be formed by neighbors maximizing the total weight of \hedges connecting them:
\begin{equation}\label{eq:coarsening_goal}
    \begin{aligned}
        Score_{G}(\gamma) &= \sum_{e \in E} \sum_{n' \in N'} \omega(e) \cdot (\abs{\{n \in n' \sothat n \in e\}} - 1) \\
        &= \sum_{e \in E} \omega(e) \cdot (\abs{e} - \abs{\{\gamma(n) \sothat n \in e\}}) \text{.}
    \end{aligned}
\end{equation}

The process now repeats analogously using $G'$ as input, and onward.
Running for a runtime-dependent $l$ levels in total, producing the $\gamma_1, \dots, \gamma_l$ sequence of coarsening maps.
Coarsening stops as soon as the lowest possible valid number of nodes $\lceil \nicefrac{\abs{N}\,}{\Omega} \rceil$ is reached.
Alternatively, it stops when no further valid clusters can be built.

Coarsening itself produces our initial partitioning, with clusters on the $l$-th level coinciding with the initial partitions.
This works by recognizing that the goal of coarsening, maximizing $Score_{G}(\gamma)$, is the dual of partitioning's minimization of $Conn_{G}(\gamma)$ when $\gamma$ is interpreted as defining partitions.
Indeed, for fixed $G$, the terms $\sum_{e \in E} \omega(e)\abs{e}$ and $\sum_{e \in E} \omega(e)$ are constant, rendering Eq.~\ref{eq:coarsening_goal} and Eq.~\ref{eq:connectivity} equivalent objectives up to an additive constant.
Moreover, with no constraints on the number of partitions, the lowest cut cost naturally occurs with an almost minimal number of partitions~\cite{AxonFlow}.

The initial $\rho$ can be recovered as $\rho(n) = \gamma^l(\gamma^{l-1}(\dots \gamma^1(n)))$, by projecting partitions backward through levels.
Coarsening, however, only aims to minimize connectivity on a per-level basis, implying that, as levels uncoarsen, gaps for improvement appear.
Hence, every level is followed by local refinement techniques \cite{hMETIS_vlsi}.

During refinement, each node is moved to a different partition if doing so happens to fully disconnect some \hedges from its current one, sparing more cuts than it creates \cite{FMpartitioning}.
In detail, a node $n \in N$ of a level's input \hgraph is moved from its partition $p_s$ to $p_d \in P$~if:
\begin{equation}\label{eq:improving_move}
    \textstyle\sum_{e \in \mathcal{I}(n) \text{ s.t. } pins(p_s, e) = 1} \omega(e) > \textstyle\sum_{e \in \mathcal{I}(n) \text{ s.t. } pins(p_d, e) = 0} \omega(e) \text{ .}
\end{equation}
Again, only enacting moves within constraints.
Refining at every level leverages the reduced graph size to propagate improvements efficiently to many original nodes.

\section{Efficient Hypergraphs Handling on GPU}\label{sec:hypergraph_handling}

\subsection{The GPU Hierarchical Parallelism Model}\label{subsec:GPU_parallelism_model}

\begin{figure*}[t]
    \centering
    \begin{minipage}[t]{0.71\textwidth} 
        \centering
        \includegraphics[width=\linewidth]{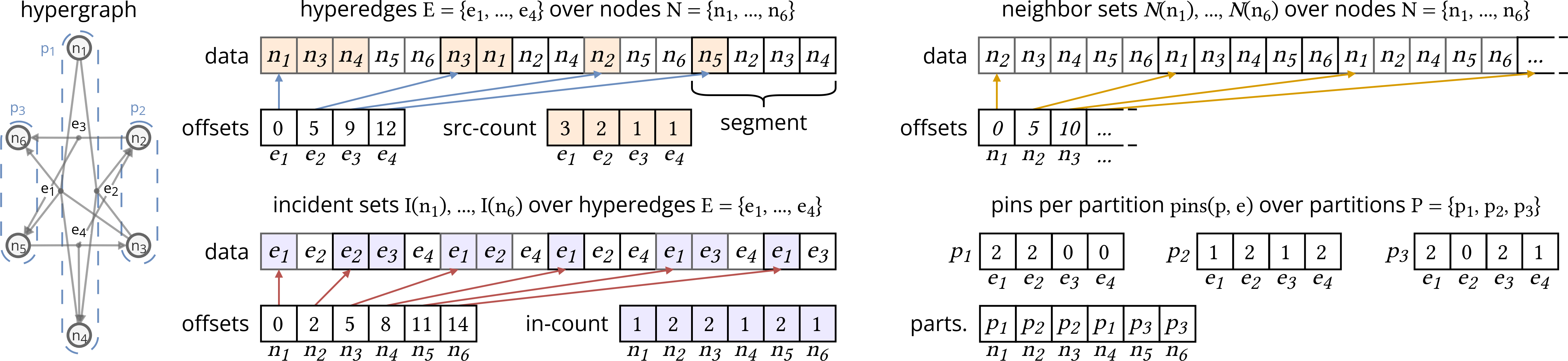}
        \captionof{figure}{Example of a hypergraph's compressed sparse memory representation.}
        \label{fig:data_structures}
    \end{minipage}
    \hfill
    \begin{minipage}[t]{0.26\textwidth} 
        \centering
        \includegraphics[width=\linewidth]{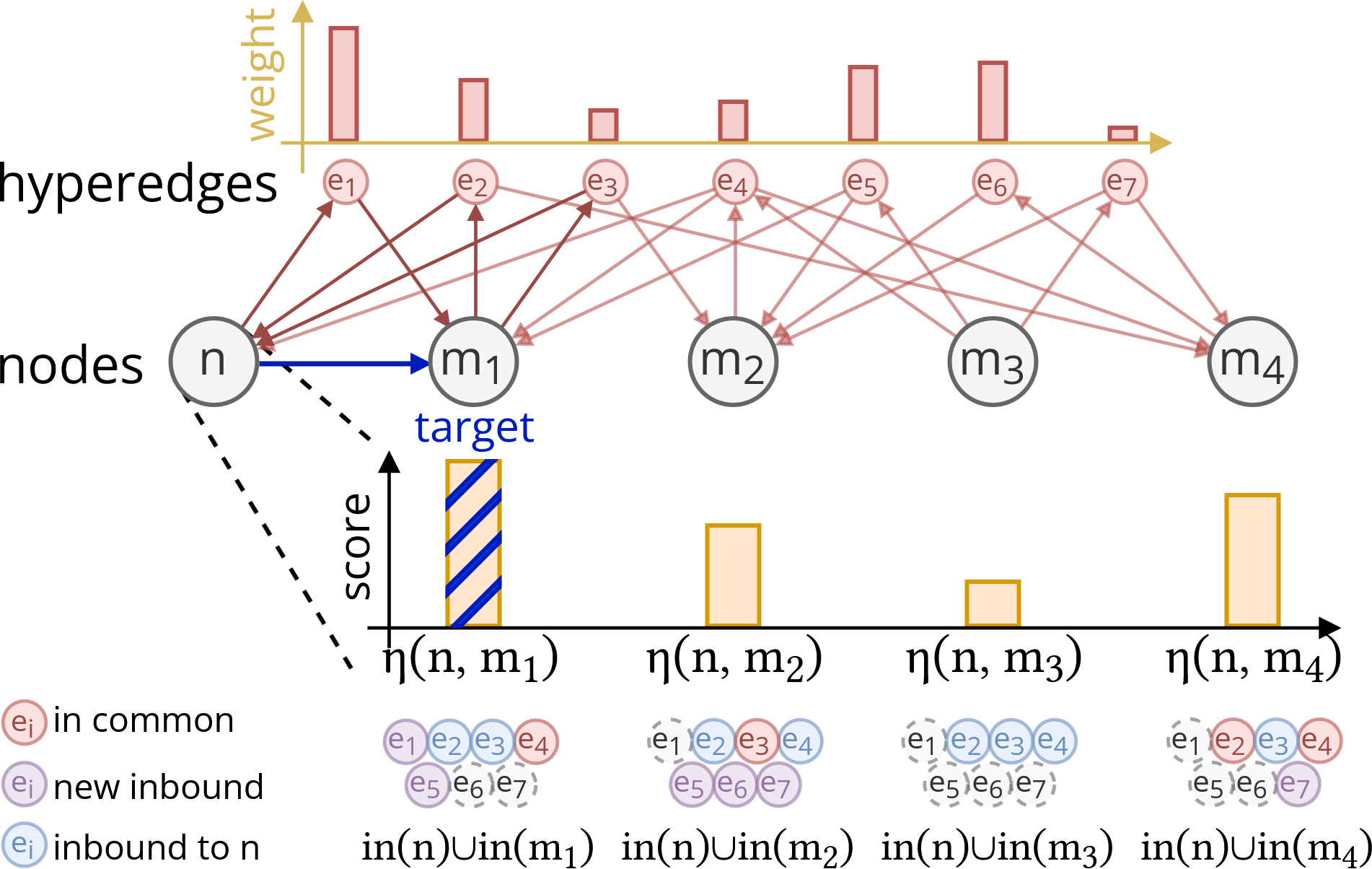}
        \hspace*{-0.05\textwidth}%
        \begin{minipage}{1.1\textwidth}
            \vspace{5.5pt} 
            \captionof{figure}{Histogram over $n$'s neighbors.}
            \label{fig:histogram}
        \end{minipage}
    \end{minipage}
    \vspace{-22pt}
\end{figure*}

Our present terminology hinges on CUDA, which is our API of choice for general-purpose processing on GPU.
From an architectural perspective, a GPU comprises several streaming multiprocessors, each handling several threads grouped in blocks.
Internally, a multiprocessor breaks a block into warps, sets of 32 threads that undergo SIMD execution.
Threads have access to a limited number of registers, while blocks can allocate a small amount of shared memory, a scratchpad seen by all their threads.
Any other data resides in global memory, backed by VRAM.
The result is a hierarchical parallelism model, spanning blocks, warps, and threads.

For a \hgraph $G(N, E, \omega)$, problem size scales with nodes $\abs{N}$, \hedges $\abs{E}$, and pins count $\sum_{e \in E}\abs{e}$.
In contrast, \hedge cardinality $d = \max_{e \in E}(\abs{e})$, node incidence degree $h = \max_{n \in N}(\abs{\mathcal{I}(n)})$, and partitions count $\abs{P}$ are local, structural parameters that remain comparatively small in practical instances.
Hereafter, we thus assume $\abs{N}, \abs{E} \gg h, d$ and parallelize first across nodes and \hedges, while confining structural iterations to inner parallelism levels.

\Hgraph algorithms are based on nested traversals.
A flat iteration over nodes or \hedges forms the outer loop.
When incidence matters, each node expands to its incident \hedges; exploring neighbors further expands each \hedge into its pins.
This progresses like:
\begin{equation}\label{eq:chained_iterations}
    \underbrace{\phantom{(}\forall \, n \in N\phantom{)}}_{\text{nodes }\rightarrow} \; , \; \underbrace{\phantom{(}\forall \, e \in \mathcal{I}(n)\phantom{)}}_{\!\!\!\!\!\!\!\!\text{incident hedges } \rightarrow\!\!\!\!\!\!\!\!} \; , \; \underbrace{\phantom{(}\forall \, m \in e\phantom{)}}_{\text{neighbors}} \qquad W = O(\abs{N}\cdot h \cdot d)
\end{equation}
where work grows multiplicatively with structural depth.

To map these traversals on GPU, we fold them twice over the parallelism hierarchy.
Outer iterations over $N$ and $E$ are distributed across blocks and warps.
A first structural iteration is handled cooperatively within a warp.
Whereas a second one is performed sequentially by entire warps -- not by threads, thus minimizing warp divergence.
Consequently, any sequential processing is restricted to the small structural dimensions $h$ or $d$ of the \hgraph.

When discussing algorithmic complexity, we report it as a pair of \textbf{work} $W$ and \textbf{span} $S$.
Respectively, the total amount of computation and the residual serial depth after mapping parallelism onto the GPU hierarchy.
Work distributed across blocks, warps, and threads is treated as ideally parallel; only operations that remain serial within a lane contribute to $S$.
For brevity, every complexity reported here is an asymptotic upper bound and we omit the $O(\cdot)$~notation.

\subsection{Hypergraph Data Structures}\label{subsec:hypergraph_data_structures}

We store \hgraphs in a compressed sparse format, with their internal structure traversed at warp granularity.
See Fig.~\ref{fig:data_structures}.
This layout both mitigates warp divergence and promotes memory access coalescing.

An \hgraph is primarily described by sets of sets, namely $E$ and $\mathcal{I}(n)$.
The memory representation of such two-level structures in compressed sparse form involves two arrays.
A segmented data array stores contiguously each linearized inner set.
Then, an array of offsets maps inner set ids to their data's starting position in the previous array.
If now one or few warps handle each segment, they will see fully coalesced accesses and little divergence.
Nodes and \hedges alike are identified by unsigned integers, their ids constituting all atoms inside sets.
With ids being a zero-based range, they double as indices in the offsets array.

To fully match the formalization in Sec.~\ref{sec:problem_definition}, \hedges and incidence sets also need their entries to be separable across $src(\cdot)$-$dst(\cdot)$ and $in(\cdot)$-$out(\cdot)$ respectively.
For this reason, we store in each \hedge segment all source pins first, while for incidence sets we store inbound \hedge ids first.
Then, we keep a secondary offsets array for each of these data structures, containing $\abs{src(\cdot)}$ and $\abs{in(\cdot)}$ respectively, to enable precise access to each subset.

These data structures reside in global memory, and their compressed layout is optimized for coalesced access by design when an entire warp is used to iterate over each segment.
It follows that warp shuffles can be used to trivialize most parallel patterns within a segment.
All kernels described in the remainder of this work implicitly rely on this layout, even if hidden behind mathematical~objects.

\section{Coarsening}\label{sec:coarsening}

\subsection{Algorithm Overview}\label{subsec:coarsening_algorithm_overview}

Coarsening starts with the construction of mutually exclusive pairs of nodes, which is carried out in two steps.
First, each node selects among its neighbors the most suitable pairing target.
Every node and its target form a candidate pair for coarsening.
Then, actual coarse nodes are determined by a maximum-weight matching computed over candidate pairs.

A node $n$'s pairing target is the neighbor it is connected to with the highest total weight.
A target must also be valid, as in target and node must be able to form a valid cluster within constraints.
Target selection involves visiting $n$'s incident \hedges and their pins, accumulating for each neighbor the total weight of \hedges it appears in.
Doing so for every node costs $W = \abs{N} \cdot h \cdot d$.
The result is a histogram over the node's neighbors:
\begin{equation}\label{eq:neighbors_histogram}
    \forall m \in \mathcal{N}(n), \: \eta(n, m) = \textstyle\sum_{e \in \mathcal{I}(n) \text{ s.t. } m \in e} \frac{\omega(e)}{\abs{e}} \text{ .}
\end{equation}
With $\eta : N \times N \rightarrow \mathbb{R}$ defaulting to zero for non-neighbors, and weights normalized by \hedge size to limit the influence of high-cardinality ones.
Finding the best valid target then takes repeated maximum extractions from the histogram followed by constraint checks until a valid neighbor is found.
With $target : N \rightharpoonup N$ and related $score : N \rightharpoonup \mathbb{R}$, this is:
\begin{equation}\label{eq:proposal_graph}
    \begin{aligned}
        & target(n) = \textstyle\max_{id} \, \textstyle\argmax_{m \in \mathcal{N}(n) \text{ s.t. } valid(n, m)} \, \eta(n, m) \text{ ,} \\
        & score(n) = \eta(n, target(n)) \text{ ,} \\
        & valid(n, m) \Leftrightarrow size(n) + size(m) \leq \Omega \, \wedge \abs{in(n) \cup in(m)} \leq \Delta \text{ .}
    \end{aligned}
\end{equation}
Checking cluster size is trivial, while distinct inbound \hedges require computing the union set size between a node's inbound set and each neighbor's.
Finally, $target$ defines the proposed candidate pairs, weighted by the corresponding~$score$.

Candidate pairs and scores form a directed weighted graph overlaying the \hgraph, that we call proposal graph.
Any node can partake in at most one cluster, so selecting mutually exclusive node pairs reduces to a maximum weighted matching problem over said proposal graph \cite{AcceleratedCoarseningProcedure}.
However, we observe that by virtue of every node proposing one candidate pair, the proposal graph is a pseudo-forest.
Additionally, both neighbor histograms and validity are symmetric, i.e. $\forall n, m \in N$ it holds $\eta(n, m) = \eta(m, n)$ and $valid(n, m) \Leftrightarrow valid(m, n)$.
Therefore, every edge entering a node must have score less than or equal to the edge leaving it, $\forall n, m \in N, \, target(m) = n \Rightarrow score(n) \geq score(m)$.
This implies that along each component's cycle the score is constant.
In particular, by the definition of $target$ with $\max_{id}$, all cycles have length two.
In other words, for every candidate pair, either the two nodes target each other, or one has a neighbor, not in common with the other, to which it connects with a higher score.
With this structure, matching admits an exact dynamic programming solution in $W = \abs{N}$ \cite{ParameterizedAlgorithms, TreesOnGPU}.

Once final mutually exclusive pairs are determined, they become the next level's coarse nodes.
Then, constructing the coarse \hgraph in full involves several set unions and reconstructions.
Notably, merging inbound sets between paired nodes and mapping \hedge pins from nodes to clusters.
Both operations requiring deduplication at scale to preserve their sets' nature.

\subsection{Materializing Neighbors}\label{subsec:materializing_neighbors}


Coarsening requires a view of each node's unique neighbors to build the $\eta$ histogram over them.
However, without knowing unique neighbors a priori, a direct implementation would require histogram bins to be overallocated, as they must also serve for deduplication.
Each node's histogram would then demand enough memory to fit up to $d \cdot h$ neighbors, an amount that easily exceeds shared memory capacity and forces histograms to spill to global memory.
Hence, histogram construction can turn into a very costly operation, requiring many random, atomic global accesses.

To minimize the cost and memory footprint of working with unique neighbors, we fully materialize $\mathcal{N(\cdot)}$ in memory once for the initial \hgraph and progressively update it in-place while coarsening.

This does not fundamentally alter the asymptotic complexity of building the histogram, a full traversal of incident \hedges and pins per node is still required.
However, it brings several advantages.
It offsets the repeated cost of deduplication from candidate pairs proposal to a single, upfront construction.
When moving down one level, coarse neighbors are computed from existing ones, progressively dealing with fewer duplicates.
Deduplicating sets of neighbors alone, with no accumulated histogram weights on them, occupies exactly half the memory.
As a result, the one-time overhead of initial neighbors construction is amortized over all its fast updates on subsequent coarsening levels.

Materialized neighbors too follow the compressed sparse format from Sec.~\ref{subsec:hypergraph_data_structures}.
Besides, unique neighbors are solely needed during coarsening, and their memory can be reclaimed afterwards.

\subsection{Candidate Pairs Proposal}\label{subsec:candidates_pair_proposal}

To construct candidate pairs, each node $n$ shall build the histogram $\eta(n, \cdot)$ over its neighbors.
With $n$'s unique neighbors now known, we load a fixed-size batch of them at once such that the histogram fits in shared memory.
A pass over incident \hedges and pins then fills the histogram, from which the highest weight valid neighbor is extracted.
The process is repeated for all neighbor batches.

We assign a warp per node $n \in N$ and have its threads visit all pins of its incident \hedges.
First, the warp collectively loads a batch of $m \in \mathcal{N}(n)$ in shared memory, then sorting it in histogram bins by id.
Each bin accumulates $\eta(n, m)$, starting from zero.
The warp proceeds to iterate over $\mathcal{I}(n)$, having threads read consecutive pins and incrementing their histogram entry by each \hedges's weight.
For every pin, a binary search is used to find its bin, and since a node can appear at most once per \hedge, increments can be non-atomic.
Subsequently, the warp sorts the histogram $\eta(n, \cdot)$, using neighbor ids as a deterministic tie-breaker.
The maximum of the batch is then repeatedly extracted, and upon passing constraints checks, it can update the highest-scoring~target.

Constraints checks on cluster size are easily performed by keeping track of $size(\cdot)$.
Instead, for inbound set sizes, we compute $\forall m \in N,\: inter(n, m) = \abs{\{e \in E \sothat n, m \in dst(e)\}}$ as the size of the intersection of $n$ and $m$'s inbound sets.
Knowing $inter(n, m)$, we can infer $\abs{in(n) \cup in(m)}$ as $\abs{in(n)} + \abs{in(m)} - inter(n, m) \leq \Delta$, enabling immediate inbound connections constraint check.
Crucially, $inter$ can be computed for almost free, since it requires a visit of every neighbor via all \hedges connecting it to $n$, exactly what is already being performed for the histogram.
In each histogram bin, say of neighbor $m$, alongside $\eta(n, m)$, we also keep a counter representing $inter(n, m)$, initially zero.
With pins of an \hedge being unique by definition, every time a neighbor is seen as the destination -- $m \in dst(e)$ -- of an \hedge in $e \in in(n)$, its counter increments.
As a result, constraint checks introduce minimal overhead to candidate pairs proposal.
See Fig.~\ref{fig:histogram} for an example.
The final kernel's span is $S = h$.

To improve candidates quality and ensure a steady coarsening process, we augment the above histogram kernel with three mechanisms.

As the \hgraph is coarsened, it often occurs that a group of multiple nodes partakes in exactly the same \hedges, thus all achieving the same score across their histograms.
With ties broken by id, every node will thus target the same lowest-id neighbor.
But this is undesirable, since all nodes end up forming a star around a single candidate pair, wasting all other equally high-score connections they shared.
In response, we introduce some deterministic noise to subtly diversify scores.
Each histogram bin is given a small pseudo-random value: $\eta(n, m) = \dots + rng(min(n, m), max(n, m))$.
With the noise being symmetric and conditioned on both nodes, histogram symmetry is preserved.
Noise caps at $10\%$ of mean \hedge weight.

A further optimization permanently purges invalid neighbors from their sets.
When checking candidate constraints, a dedicated flag is set for each invalid neighbor.
During coarse \hgraph construction, flags are preserved across set merges, marking all occurrences of the same neighbor.
Before finalizing the new coarse sets, flagged entities are filtered out.

At last, to ensure that coarsening always reaches close to the limit of constraints, we attempt a best-effort pairing of nodes that are left with no neighbors.
After the above proposal completes, all such nodes are gathered and sorted by size.
Each node, handled by a thread, then runs a binary search for its current cluster size slack and tries to atomically claim the first valid node it finds.
Contentions are broken by id.
At this stage, inbound set union sizes are overestimated with the sum of each node's inbound set size.

\subsection{Maximum Weight Matching}\label{subsec:weighted_matching}


Now that we have the proposal graph, we must isolate a subset of mutually exclusive pairs of highest total score.
This is a weighted matching problem over the pseudo-forest of candidate pairs.
Given the proposal graph's structure discussed in Sec.~\ref{subsec:coarsening_algorithm_overview}, we propose the following dynamic programming formulation of the problem.

For convenience, let us define $child : N \rightarrow \mathcal{P}(N)$ as $child(n) = \{c \in N \sothat target(c) = n\}$, giving the set of nodes targeting another.
From it, the proposal graph's invariant can be rewritten as $\forall n \in N , \, \forall c \in child(n), \, score(n) \geq score(c)$.

The basic problem addressed by this formulation is the maximum-weight matching over a subtree rooted in an arbitrary node $n \in N$.
To solve it, we introduce two values associated with each node:
\begin{itemize}
    \item $ss_{0}(n)$: maximum score of a matching in the subtree rooted in $n$ when $n$ is not matched to $target(n)$;
    \item $ss_{1}(n)$: maximum score of a matching in the subtree rooted in $n$ when $n$ is matched to $target(n)$;
\end{itemize}
For the ensuing discussion, let us isolate root nodes of the proposal graph as $R =\{n \in N \sothat target(target(n)) = n\}$, thus avoiding circular dependencies in the recursion.
Then, for every $n \in N \setminus R$, the second value can be expressed as:
\begin{equation}\label{eq:ss1}
    ss_1(n) = score(n) + \textstyle\sum_{c \in child(n) \setminus R} ss_0(c) \text{ .}
\end{equation}
While on root nodes $n \in R$:
\begin{equation}\label{eq:ss1_roots}
    ss_1(n) = score(n) + \textstyle\sum_{c \in \left ( child(n) \cup child(target(n)) \right ) \setminus R} ss_0(c) \text{ .}
\end{equation}
When node $n$ matches with its children $c'$, the first value is:
\begin{equation}\label{eq:ss0_base}
    ss_0(n) = \textstyle\sum_{c \in child(n) \setminus R, \, c \neq c'} ss_0(c) + ss_1(c') \text{ .}
\end{equation}
Thus, let $ss_{1-0}(n) = ss_1(n) - ss_0(n)$ be the return for any node $n \in N$ matching with its target.
In general, the first value will depend on a node's highest $ss_{1-0}$ children, that is:
\begin{equation}\label{eq:ss0}
    ss_0(n) = \textstyle\sum_{c \in child(n) \setminus R} ss_0(c) + \max\! \left ( 0, \, \textstyle\max_{c \in child(n) \setminus R} ss_{1-0}(c) \right )\text{ .}
\end{equation}

Let $match : N \rightharpoonup N$ define the final matching partner for each node, being undefined if the node is unmatched.
We infer $match$ from $ss_0$ and $ss_1$, starting from the case of roots $n \in R$, as:
\begin{equation}\label{eq:match_root}
    match(n) =
    \begin{cases}
        target(n) & \hspace{-3em} \text{if } ss_1(n) > ss_0(n) + ss_0(target(n)) \text{ ,} \\
        \displaystyle\argmax_{c \in child(n)} ss_{1-0}(c) & \text{otherwise} \text{ .}
    \end{cases}
\end{equation}
Finally, for every other node $n \in N \setminus R$:
\begin{equation}\label{eq:match}
    match(n) =
    \begin{cases}
        target(n) & \text{if } match(target(n)) = n \text{ ,} \\[4pt]
        \elevate[6pt]{\displaystyle\argmax_{c \in child(n)} ss_{1-0}(c)} & 
            \begin{aligned}
                & \text{if } match(target(n)) \neq n \, \wedge \\[-4pt]
                & \textstyle\max_{c \in child(n)} ss_{1-0}(c) > 0 \text{ ,}
            \end{aligned} \\[4pt]
        \text{undefined} & \text{otherwise .}
    \end{cases}
\end{equation}
This formulation exhibits optimal substructure: the optimal matching for the subtree rooted at $n$ depends exclusively on the optimal matchings of its children's subtrees.

Both values $ss_0$ and $ss_1$ can be computed for all nodes by a single bottom-up traversal of each pseudo-tree.
At branching nodes, contributions from all children accumulate, and the locally optimal child choice is refined whenever a higher-scoring subtree is encountered.
Once all upward computations are complete, a top-down traversal of each connected component can finalize the matching.

\begin{figure}[t]
    \centering
    \includegraphics[width=1.0\columnwidth]{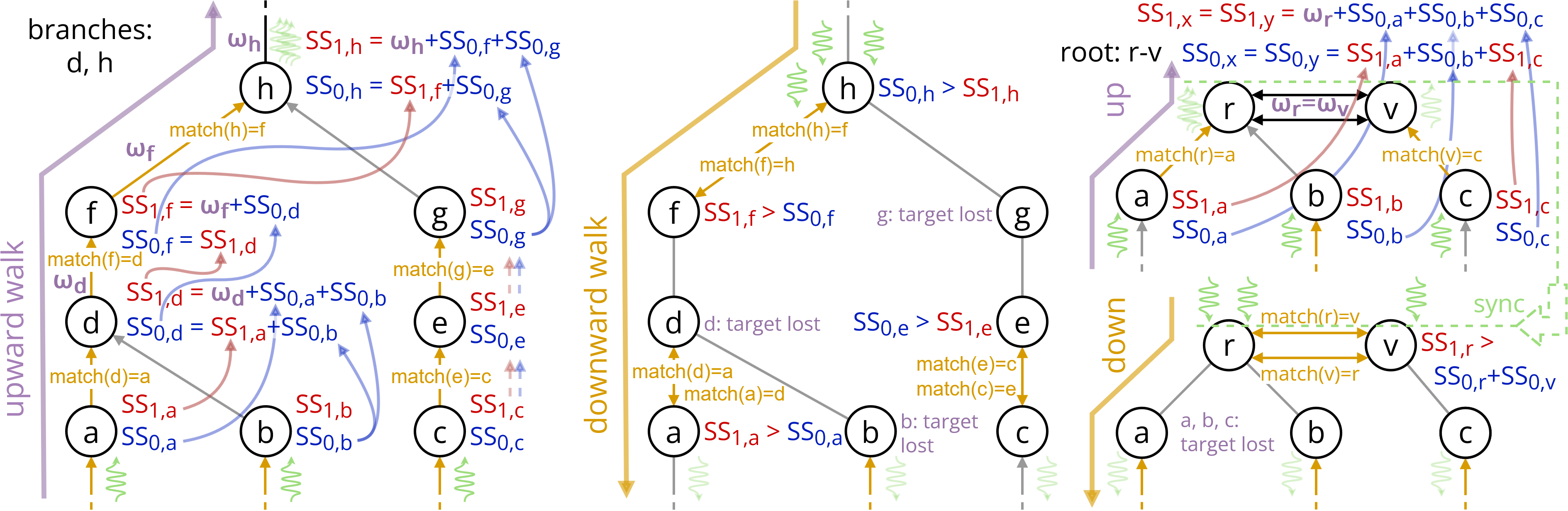}
    \vspace{-12pt}
    \caption{Dynamic programming formulation for maximum weighted matching in a two-cycle pseudo-forest.}
    \label{fig:dynamic_prog}
\end{figure}


The aforementioned bottom-up and top-down traversals can be realized in parallel by launching one thread for every leaf in the pseudo-forest.
Each thread then moves upward along the path $n, \, target(n), \, target(target(n)), \, \dots$, amending every $ss_0(n)$ and $ss_1(n)$ with the scores of the subtree it has traversed so far, see Fig.~\ref{fig:dynamic_prog}.
To support the upward walk, we provisionally build $match$ asymmetrically: $match(n)$ is continuously updated to point to the best current $child(n)$ and tracked as part of a tuple ($ss_{1-0}(match(n))$,~$match(n)$).
Before moving upward, a thread on node $n$ tries to use $ss_{1-0}(n)$ to claim $match(target(n))$.
Claiming occurs via an atomic lexicographic max over the target's tuple with ($ss_{1-0}(n)$,~$n$).
This is consistent with the tie-breaking used for candidate pairs, ensuring the deterministic convergence of $match$.
Continuing, $ss_0(n)$ is atomically added to $ss_1(target(n))$, and $ss_0(target(n))$ is recomputed based on the claim's outcome.

Once every thread reaches a root, they all synchronize, settle the matching for the roots, and the downward walk begins.
Each thread retraces its full path backwards, checking if its claims were successful, and if so, permanently constructing matches.
A node $n$ is matched with its target if, at that point, it still holds $match(target(n)) = n$.
If so, $n$ discards whichever claims it received and imposes $match(n) = target(n)$.
Otherwise, the same reasoning is repeated one step down the line: $n$ remains unmatched to its target and will end up matched with the current $match(n)$, i.e. the highest $ss_1$ holder among $child(n)$.

While some threads may partially walk the same path, that doesn't constitute a problem, as they will take the same decisions.
Rather, to finalize the exclusive part of its path, each thread must know everything that happened between its root and leaf, therefore some redundant work is inevitable to run the descent in parallel.
This routine's span is the maximum height of a tree, that typically being a very small value, we treat as a constant, hence $S = 1$.

To increase the likelihood of nodes forming clusters, candidate pairs proposal (Sec.~\ref{subsec:candidates_pair_proposal}) actually produces $\Pi$ candidates, in order of decreasing score.
These correspond to the top-$\Pi$ valid neighbors per histogram, leading to $\Pi$ proposal graphs.
Matching then repeats for $\Pi$ rounds; $\Pi = 4$ by default.
To preserve the invariant of each proposal graph, nodes matched in earlier rounds are removed from all subsequent graphs.
Any instance of $target$ pointing towards one such node thus reverts to being undefined.
Ultimately, the combination of deterministic noise and multiple candidates ensures that almost every node finds a pair.



\subsection{Coarse Hypergraph Construction}\label{subsec:coarse_hypergraph_construction}

\begin{figure*}[t]
    \centering
    \begin{minipage}[t]{0.34\textwidth} 
        \centering
        \includegraphics[width=\linewidth]{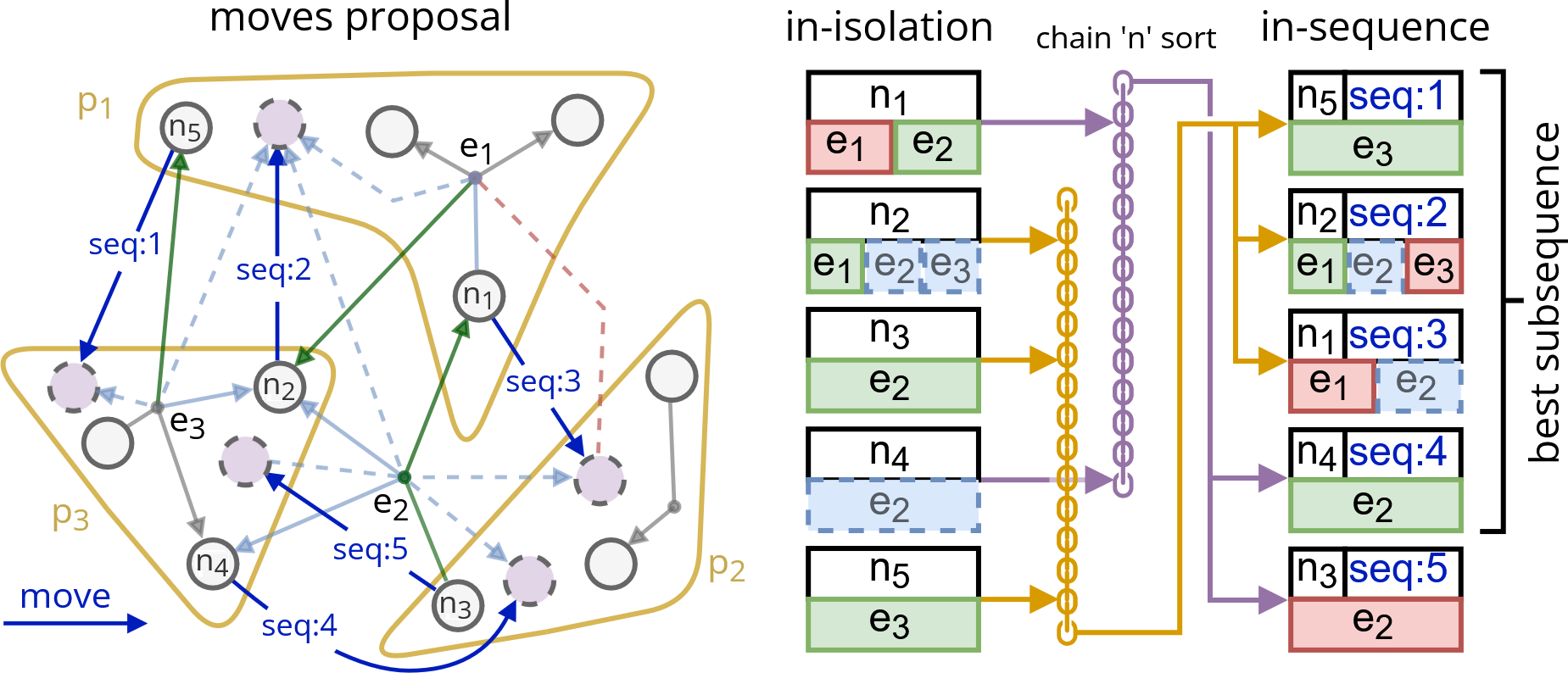}
        \hspace*{-0.04\textwidth}%
        \begin{minipage}{1.05\textwidth}
            \vspace{5.5pt}
            \captionof{figure}{Moves sequence-of-chains construction.}
            \label{fig:moves_sequence}
        \end{minipage}
    \end{minipage}
    \hfill
    \begin{minipage}[t]{0.62\textwidth} 
        \centering
        \includegraphics[width=\linewidth]{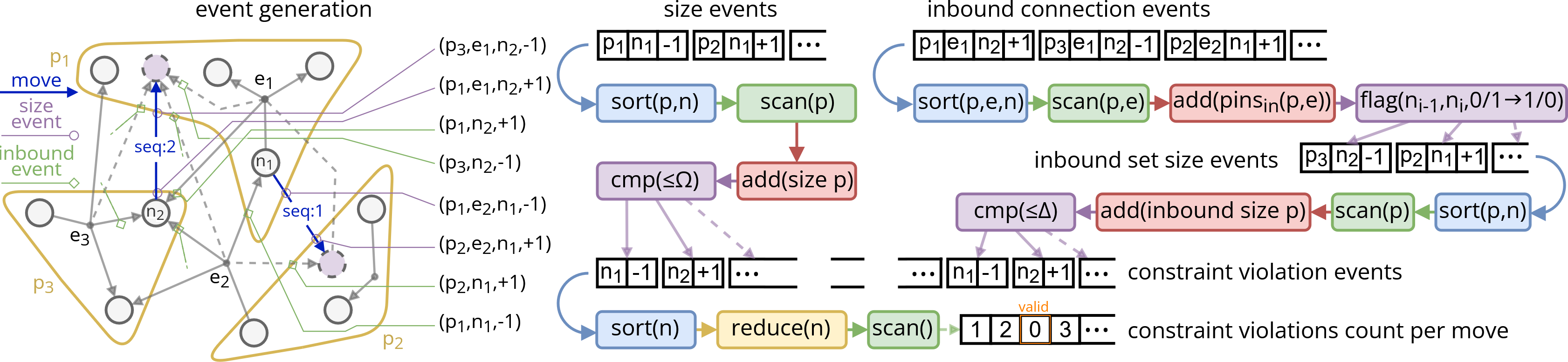}
        \captionof{figure}{Events-based moves validity check.}
        \label{fig:events_generation}
    \end{minipage}
    \vspace{-22pt}
\end{figure*}

With $match$ defining clusters, coarsening finishes by constructing the coarse \hgraph $G'(N', E', \omega')$ as per Sec.~\ref{sec:multi_level_scheme}.
For that, the nodes-to-coarse-nodes map $\gamma$ is built as $\forall n \in N, \, \gamma(n) = \{n, match(n)\}$ if $match(n)$ exists and $match(match(n)) = n$ else $\gamma(n) = \{n\}$.
Then, constructing coarse instances of all two-level structures, $E$, $\mathcal{I}(\cdot)$, $\mathcal{N}(\cdot)$ requires a sequence of map patterns through $\gamma$ and~set~unions.

Both operations involve deduplication and the final set size isn't known a priori, therefore a compressed representation can't be built directly.
Constructing many sets in parallel requires two phases: first, an oversized data array is built and deduplication takes place over it, then, its segments are packed in the final compressed form.

Each coarse set is assigned to a block and given an oversized global memory segment to work with.
The block is also given the maximum possible amount of shared memory; both memories are configured as closed-hashing hash-sets.
Threads in the block collectively read elements to insert in their set, using $\gamma$ as needed to map nodes to clusters before inserting them.
Insertion of each element is first attempted in shared memory, going to global memory only upon a successful insertion or exhausted probe-length.
Once all elements have been seen, the content of shared memory is cooperatively moved to the remaining holes in global memory.
While this unfolds, each set's size is tracked, with a subsequent prefix-sum of set sizes providing both the final compressed data array size and the offset for each set.
Lastly, a packing operation scatters each oversized array segment into its final segment in the newly allocated compressed~data~array.

The major remaining issue is knowing how large of an allocation is needed for each oversized array segment.
\Hedges do not involve any set union, hence each current segment size is a viable upper bound.
For neighbors, instead, each new segment size must be the sum of the sizes of segments that will be merged into it.
Finally, for incidence sets there is a shortcut: with \hedges coarsened first, count how many times each coarse node occurs as a new pin; that count is the number of distinct \hedges incident to every~coarse~node.

Note that to ensure the correctness of inbound sets while guaranteeing no self-cycles, eventual duplicates between $src(\cdot)$ and $dst(\cdot)$ or $in(\cdot)$ and $out(\cdot)$ are discarded from $src(\cdot)$ and $out(\cdot)$, respectively.

\section{Uncoarsening and Refinement}\label{sec:refinement}

\subsection{Algorithm Overview}\label{subsec:refinement_algorithm_overview}

Local refinement improves a partitioning by selectively moving nodes across partitions.
Choosing suitable moves takes two steps.
First, each node independently selects the partition it would rather belong to, proposing a move.
Then, a subset of moves is found such that, when applied together, they lead to a valid state of lowest possible connectivity.

Each node $n$ proposes its move in-isolation, only with regard to the current partitioning.
By Eq.~\ref{eq:connectivity}, a cut is only avoided when an \hedge has no pins left in a partition.
So, a favorable move is one that fully disconnects \hedges from $n$'s current partition while introducing cheaper connections, if any, to its new partition, as per Eq.~\ref{eq:improving_move}.
To find such moves, a node must count the number of pins each of its incident \hedges owns in every partition, what we defined as $pins(p, e)$ in Sec.~\ref{subsec:objective_and_constraints}.
Leaving the current partition spares the weight of \hedges with only a single pin, the node, remaining in it.
Entering another partition costs the total weight of \hedges that currently hold no pins in it.
The difference between two said quantities is the gain in connectivity for a move:
\begin{equation}\label{eq:refinement_move_selection}
    \begin{aligned}
        & saving(n) = \textstyle\sum_{e \in \mathcal{I}(n) \text{ s.t. } pins(\rho(n), e) =\;\! 1} \omega(e) \text{ ,} \\
        & loss(n, p) = \textstyle\sum_{e \in \mathcal{I}(n) \text{ s.t. } pins(p, e) =\;\! 0} \omega(e) \text{ ,} \\
        & gain(n, p) = saving(n) - loss(n, p) \text{ .}
    \end{aligned}
\end{equation}
Moving node $n$ to partition $p$ is favorable if $gain(n, p) > 0$.
In fact, Eq.~\ref{eq:refinement_move_selection} is a direct adaptation of Eq.~\ref{eq:improving_move}.
The move with highest gain is proposed by the node as $move : N \rightarrow P$, $move(n) = \max_{id} \argmax_{p \in P} gain(n, p)$.
Accordingly, let $p_s^n \xrightarrow{n} p_d^n$ denote a move of node $n$ from partition $p_s^n = \rho(n)$ to $p_d^n = move(n)$.
Ultimately, computing gains calls for another neighborhood traversal, with $W = \abs{N} \cdot h \cdot d + \abs{N} \cdot \abs{P}$.


So-obtained moves have been proposed separately, but to maximize gain, several of them shall be applied at once.
Deciding which moves to apply equates to finding the subset of moves collectively leading to a valid maximum gain partitioning.
However, moves easily interfere, influencing each other's gain and feasibility, making this a problem only solvable in exponential time.
For this reason, we revise a heuristic from \cite{HyperG}.
We sort moves into a sequence and only apply a subsequence of them from the first one onward.
In building the sequence, we assemble chains of moves with source-to-destination concatenation and involving nodes of similar size.
Chains are then sorted by total gain to form a sequence that promotes feasible and favorable swaps and cyclic exchanges.

With moves ordered, the problem of deciding which moves to apply reduces to finding the longest subsequence of improving moves landing on a valid state.
To solve it, each move recomputes its gain in-sequence, i.e. assuming all moves before it already applied.
Analogously, validity over constraints is computed as of every move along the sequence.
A filtered maximum extraction then leads to the desired subsequence, whose moves are applied.
The whole refinement process repeats $\Theta$ times per level; $\Theta = 16$ by default.

\subsection{Refinement Moves Proposal}\label{subsec:refinement_moves_proposal}

Proposing moves in-isolation starts with each node computing gains over partitions.
Every warp handles a node $n$, currently in partition $p_s$.
First, it allocates one variable for every partition in shared memory, representing $saving(n) = 0$ and $\forall p \in P \setminus \{p_s\}, \, loss(n, p) = 0$.
Then, threads iterate $n$'s incident \hedges.
For every \hedge, if $pins(p_s, e) = 1$, $saving(n)$ increments by $\omega(e)$, and for every partition $p_d \neq p_s$, if $pins(p_d, e) = 0$ , $loss(n, p_d)$ increments by $\omega(e)$.
A map-reduce operation computes each $gain$ from $loss$ and $saving$, finds the maximum, and yields the node's proposed move.

Gains exclusively depend on the count of \hedge pins per partition, $pins(\cdot, \cdot)$.
With the same values of $pins$ reused multiple times across nodes, they shall be precomputed in parallel \cite{HyperG}.
Thus, we prepare a matrix $\abs{P} \times \abs{E}$ with the values of $pins(p, e)$ before commencing refinement, see Fig.~\ref{fig:data_structures}.
Every warp is assigned an \hedge and allocates one counter in shared memory per partition.
Its threads then iterate over the \hedge's pins and atomically increment counters after applying $\rho$ to every read pin.
Overall, this yields a span $S = 1$.

At this stage, the first half of refinement repetitions can propose moves that invalidate a partition by size.
The second half, instead, imposes $size(n) + \sum_{m \in p_d} size(m) \leq \Omega$ to focus on smaller, consistent improvements.
In any case, the next steps strictly enforce final validity.

\subsection{Moves Sequence Construction}\label{subsec:moves_sequence_construction}

The efficacy of refinement is strongly limited by the joint feasibility of multiple moves.
For several moves to consistently return to a valid state, they shall realize swaps or cyclic exchanges between nodes with compatible effects on the constraints of their partitions.
Thus, the goal for sequence construction is to produce a total ordering of moves that consists of several chains -- paths or cycles -- preferably of length two or more.
A chain concatenates moves each departing from the destination of the previous one, with minimal size and inbound set variation between their affected nodes.
Arranging chains by decreasing internal total gain gives the final sequence.
Formally, this is a weighted path covering problem that we solve greedily.


Moves are first sorted by the ordered pair $(p_s^n, -gain(n, p_d^n))$, then chains are constructed over several rounds.
Each move is assigned a predecessor, initially empty.
In a round, each free end of a chain (initially every move), consider it node $n$'s move, sifts a window of candidate successors $m \in N$ among moves with $p_s^m = p_d^n$, selecting the one maximizing a compatibility grade $gain(m, p_d^m) - \alpha \cdot \abs{size(n) - size(m)} - \beta \cdot \abs{\abs{in(n)} - \abs{in(m)}}$; for us $\alpha = {\scriptstyle 10}^{-6}$, $\beta = {\scriptstyle 10}^{-7}$, and window size is $256$.
Conflicts -- multiple chains choosing the same successor -- are resolved by parallel atomic maximization on the grade and moved node id.
Moves that secured a successor are frozen, and the process repeats up to $16$ times or until no loose ends remain.
See an example in Fig.~\ref{fig:moves_sequence}.

After all rounds complete, the successor–predecessor relation will have induced disjoint paths and cycles.
Subsequences are extracted by following predecessor links, simultaneously computing their total gain.
Chains are ranked by decreasing total gain and concatenated into the final sequence, that by the one-to-one relation between nodes and moves we formalize as a total order of nodes $\prec_{seq}N$.

With moves sorted by their in-isolation gain, their in-sequence gain must be inferred.
For every node $n$ with move $p_s^n \xrightarrow{n} p_d^n$, a warp iterates over $n$'s incident \hedges and their pins.
For every \hedge $e \in \mathcal{I}(n)$, consider only pins $m \in e$ whose moves precede $n$'s in the sequence, $m <_{seq} n$, and related moves $p_s^m \xrightarrow{m} p_d^m$.
Then, two conditions may arise on node $n$ after seeing all such $m$-s:
\begin{align}\label{eq:in_sequence_gain}
    & \begin{aligned}
        & \text{if \;} \underbrace{\abs{\{m \sothat p_d^n = p_s^m\}}}_{m\text{ leaving }p_d^n} - \underbrace{\abs{\{m \sothat p_d^n = p_d^m\}}}_{m\text{ also entering }p_d^n} = pins(p_d^n, e) > 0 \\
        & \text{or } \exists \:\! m \text{ s.t. } p_s^n = p_d^m \text{ and } pins(p_s^n, e) = 1 \\
        & \text{then } gain(n, p_d^n) \, \mathrel{-\!=} \, \omega(e) \\
    \end{aligned} \\[6pt]
    & \begin{aligned}
        & \text{if \;} \underbrace{\abs{\{m \sothat p_s^n = p_s^m\}}}_{m\text{ also leaving }p_s^n} - \underbrace{\abs{\{m \sothat p_s^n = p_d^m\}}}_{m\text{ entering }p_s^n} = pins(p_s^n, e) - 1 > 0 \hspace{-4em} \\
        & \text{or } \exists \:\! m \text{ s.t. } p_d^n = p_d^m \text{ and } pins(p_d^n, e) = 0 \\
        & \text{then } gain(n, p_d^n) \, \mathrel{+\!=} \, \omega(e)
    \end{aligned}
\end{align}
In the first case, \hedge $e$, which was not cut in isolation, becomes cut in the sequence, either due to $n$ now being the first to re-enter an earlier disconnected partition, or it no longer being the last node of $e$ to leave its current partition.
Conversely, in the second case, an additional cut on \hedge $e$ is spared because another node $m$ entered $n$'s destination for the first time before $n$, or $n$ suddenly became the last node of $e$ in its partition.
Again, by the neighborhood traversal,~$S = h$.

A sequence of nodes, chained and ranked by in-isolation gain, and carrying their in-sequence gain, is thus available.

\subsection{Events-based Constraint Checks}\label{subsec:events-based_constraint_checks}

\begin{figure*}[t]
    \centering
    \vspace{-6pt}
    \includegraphics[width=\textwidth]{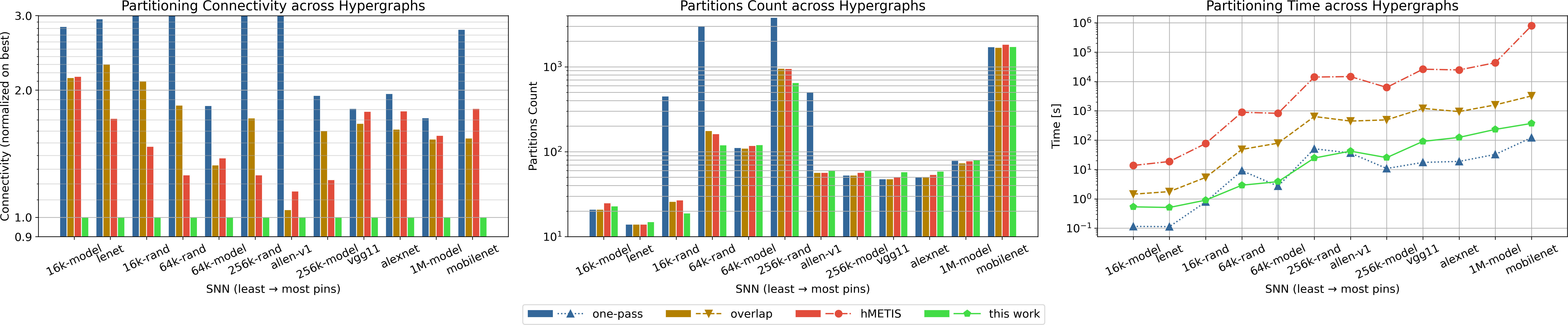}
    \vspace{-13pt} 
    \caption{Partitioning results comparison with three sequential methods across twelve SNNs hypergraphs.}
    \vspace{-8pt} 
    \label{fig:results}
\end{figure*}

\begin{figure*}[t]
    \centering
    \begin{minipage}[t]{0.3625\textwidth} 
        \centering
        \includegraphics[width=\linewidth]{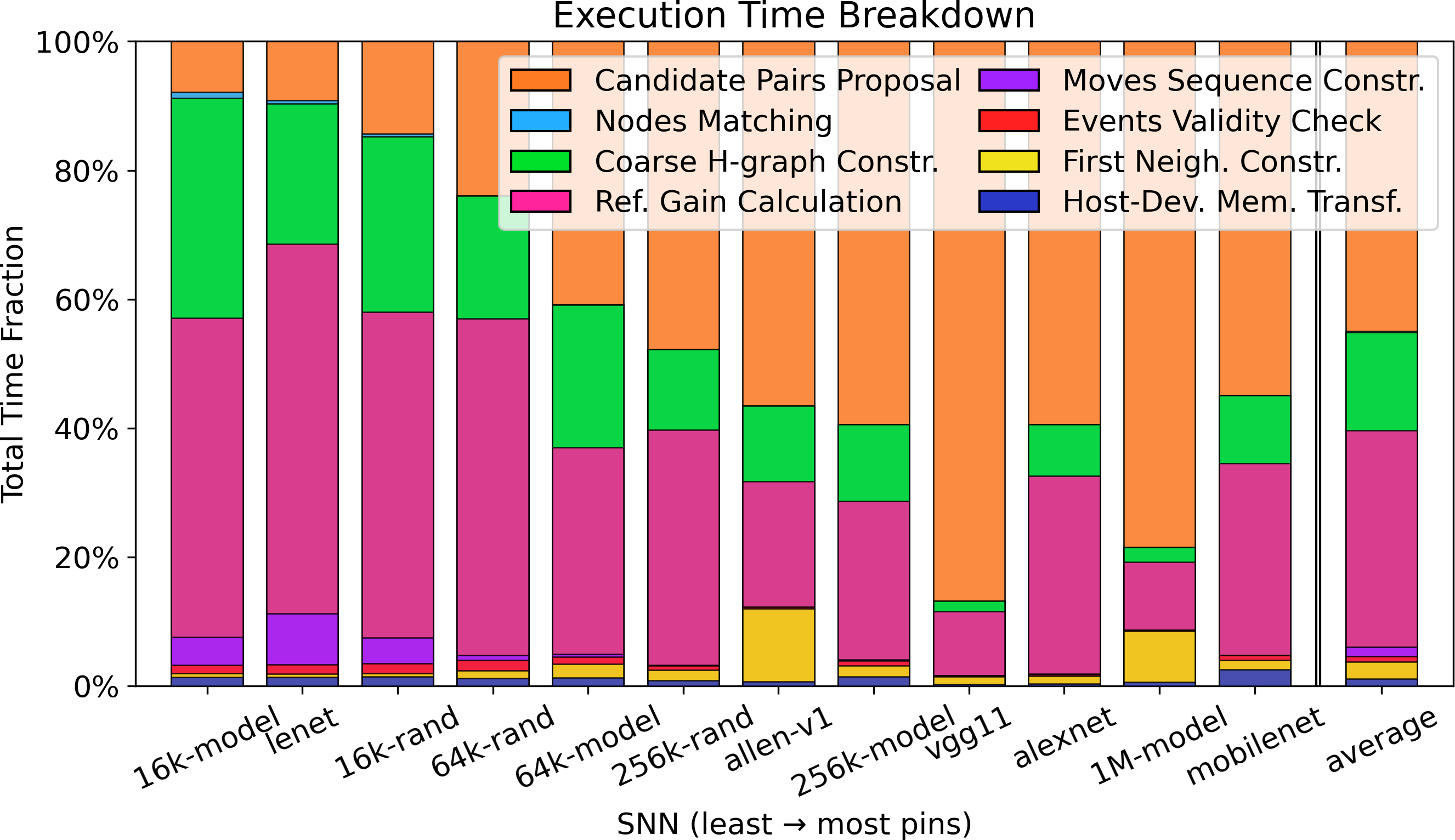}
        \vspace{-13pt}
        \captionof{figure}{Algorithm steps execution time breakdown.}
        \label{fig:exec_breakdown}
    \end{minipage}
    \hfill
    \begin{minipage}[t]{0.3625\textwidth} 
        \centering
        \includegraphics[width=\linewidth]{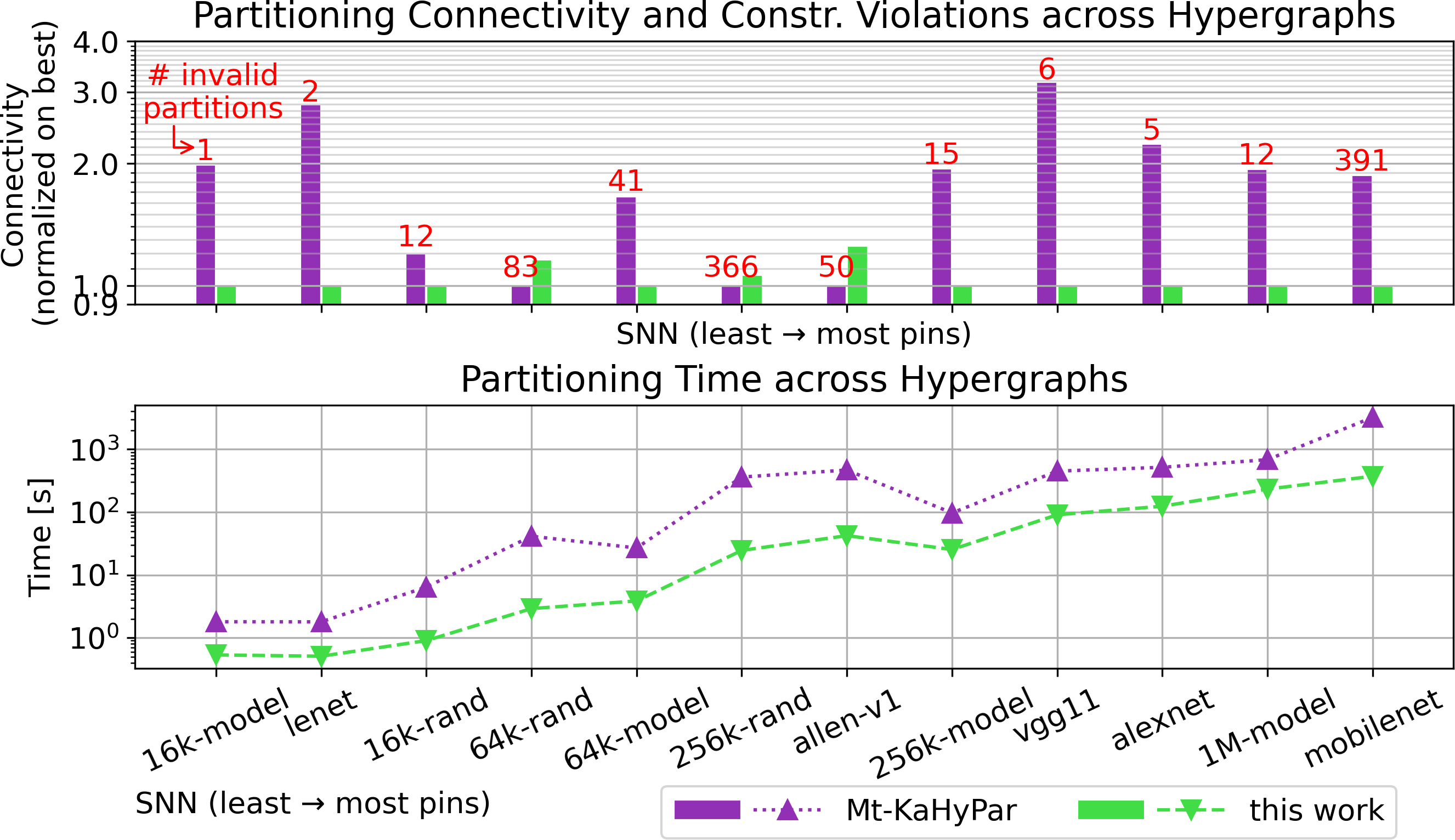}
        \vspace{-13pt}
        \captionof{figure}{Partitioning results vs Mt-KaHyPar~\cite{MtKaHyPar}.}
        \label{fig:results_parallel}
    \end{minipage}
    \hfill
    \begin{minipage}[t]{0.235\textwidth} 
        \centering
        \includegraphics[width=\linewidth]{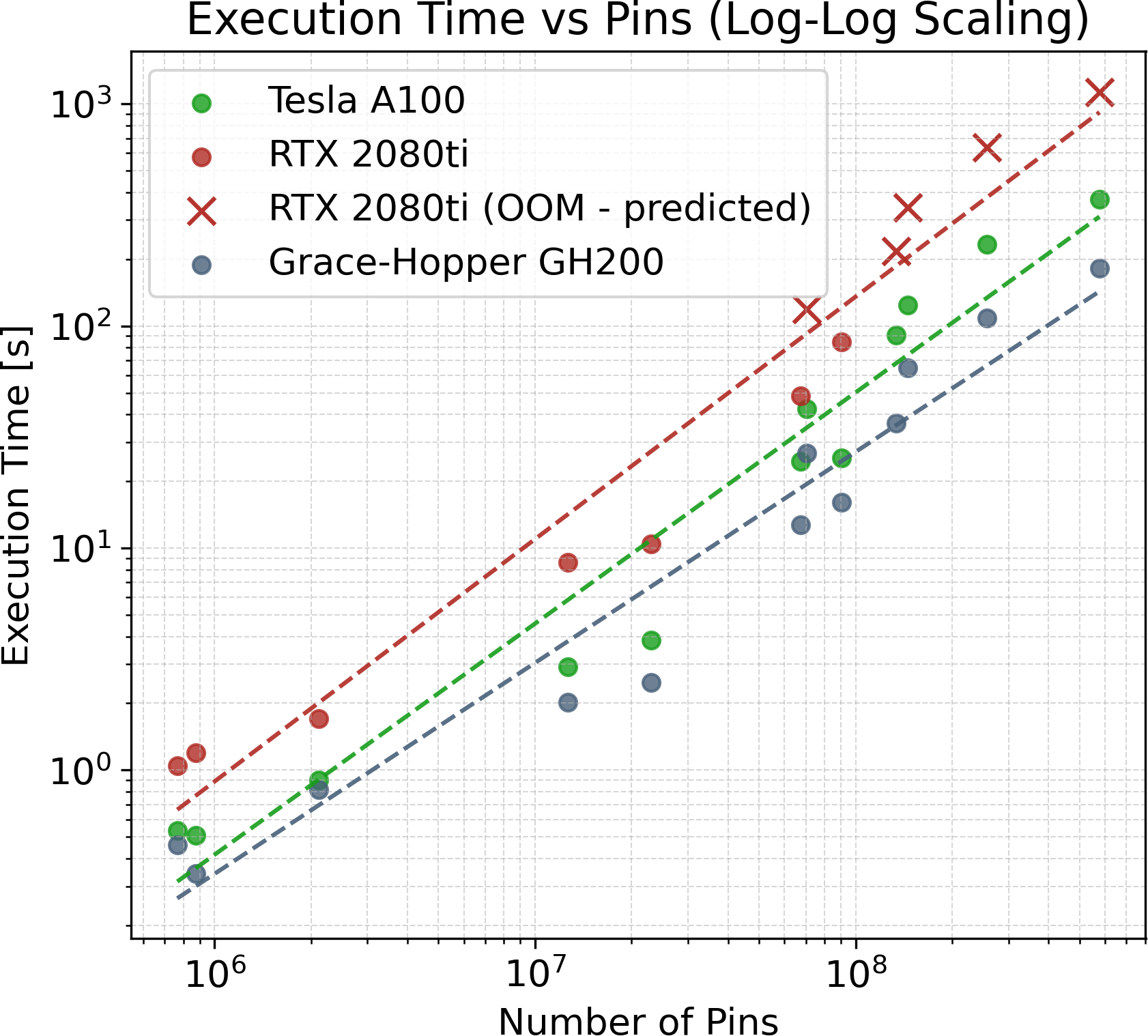}
        \vspace{-13pt}
        \captionof{figure}{Scalability across GPUs.}
        \label{fig:gpu_scaling}
    \end{minipage}
    \vspace{-14pt}
\end{figure*}


Simultaneously assessing the partitioning's validity at every move in the sequence requires reconstructing every intermediate state of all partition sizes and, critically, inbound sets.
Materializing all such states is prohibitive; consequently, we handle constraint checks sparsely through "events".
First, each move generates events for every partition size and inbound set size variation it causes, carrying the variation's delta as payload.
Next, with a series of parallel patterns over deltas, we infer each move's \hbox{validity, see Fig.~\ref{fig:events_generation}}.

Crucially, a move alters partition size for exactly two partitions, and affects inbound set sizes only through transitions of $pins_{in}(p, e)$ across zero.
Exploiting this structure, every move of a node $n$, $p_s^n \xrightarrow{n} p_d^n$, generates two size-event \mbox{3-tuples} ($p_s^n$, $n$, $-size(n)$), ($p_d^n$, $n$, $+size(n)$), along with several inbound connection events given by the \mbox{4-tuples} $\forall e \in in(n)$, \mbox{($p_s^n$, $e$, $n$, $-1$)}, \mbox{($p_d^n$, $e$, $n$, $+1$)}.

Each size event ($p$, $n$, $\delta$) means that partition $p$'s size changes by $\delta$ with $n$'s move.
After being sorted by ($p$, $n_{seq}$), their deltas are prefix summed for each $p$; thus, each event stores its partition's cumulative size variation up to its move along the sequence.

For the inbound \hedges count, we here rely on $pins_{in}(p, e)$, rather than a set-based formulation, as shown in Sec.~\ref{subsec:objective_and_constraints}.
Therefore, we modify the precomputed $pins$ matrix to represent $pins_{in}$ with a quick pass that subtracts the count of outbound \hedges.

Each inbound set event ($p$, $e$, $n$, $\delta$) means that $pins_{in}(p, e)$ changed by $\delta$ after $n$'s move.
They are first sorted using ($p$, $e$, $n_{seq}$) and prefix summed using ($p$, $e$) as keys.
To track inbound set size changes, $pins_{in}(p, e)$ is added to each respective delta, giving $e$'s running inbound pin count on $p$.
A new event \mbox{($p$, $n$, $-1$)} is then emitted whenever such counts transition from $1$ on the event before to $0$, or ($p$, $n$, $+1$) when turning from $0$ to $1$.
Resulting events are again sorted by key ($p$, $n_{seq}$) and prefix summed per $p$, giving each $p$'s distinct inbound \hedges count variation as of $n$'s move~in~the~sequence.

With all events sorted by ($p$, $n_{seq}$), adding initial set sizes $\abs{p}$ and $\abs{\{e \in E \sothat pins_{in}(p, e) > 0\}}$ respectively yields used constraint capacities move by move.
A subsequent comparison with $\Omega$ or $\Delta$ shows if $p$ is valid after moving $n$.
Then, looking at pairs of events for consecutive moves of $m <_{seq} n$ in the sequence tells if moving $n$ was responsible for invalidating or re-validating partition $p$ based on how it was left by $m$.
This spawns a final sequence of events for every time a partition changes to invalid ($n$, $+1$) or valid ($n$, $-1$).
When sorted and reduced by $n_{seq}$, their prefix sum is the count of active constraint violations after each move.
Only moves with count zero are valid, finding the one of maximum cumulative gain gives the subsequence of moves to apply.

\section{Experimental Results}\label{sec:experimental_results}

\subsection{Experimental Setup}\label{subsec:experimental_setup}

\slantedssnstableall

To test our solution, we conduct two types of experiments.
An application-driven evaluation under our constraints, followed by validation under standard benchmarks and metrics.

First, we partition 12 \hgraphs originating from SNNs and their mapping constraints on neuromorphic hardware \cite{BenchmarkSNNs}, see Tab.~\ref{tab:snns_all}, chosen for their diversity in size and topology.
Networks rapidly grow in size, from 1\MM to over 500\MM pins.
Topologies, meanwhile, vary from a mostly local and regular structure on \texttt{-model} networks, to a small-world, erratic structure on \texttt{-rand} ones.
Hence, the latter exhibit significantly larger neighborhoods.
In particular, we present detailed performance metrics for two \hgraphs, the \texttt{256k-model}, a VGG-like ANN \cite{VGG} converted to SNN, and the \texttt{allen-v1} \cite{AllenV1} neural model.
Together, they are representative of both extremes in \hgraph topologies, from regular to highly not so, and all our other benchmarks follow either of their trends.

Our baseline comprises three heuristics running sequentially on CPU.
An implementation of the multi-level scheme in hMETIS adapted to our constraints \cite{hMETIS_k_way, AxonFlow}.
And two simple algorithms originating from SNN mapping tools. 
A greedy "overlap" \cite{AxonFlow} heuristic that co-locates nodes based on their incidence sets' overlap.
And a trivial "one-pass" \cite{MappingVeryLargeSNNtoNHW} algorithm that fills one partition after the other during a single pass over nodes, driven solely by constraints.
In addition, we compare against the SoTA Mt-KaHyPar~\cite{MtKaHyPar} multi-threaded CPU partitioner.
However, Mt-KaHyPar lacks support for incidence constraints, therefore serving as a qualitative reference.

\complexitytable

Using the same setup, we also perform a set of ablation studies to measure the impact of our design choices, namely the materialization of neighbors, optimal matching, and refinement moves chaining.
We also study the effect of varying our parameters for the number of coarsening candidates ($\Pi$) and refinement repetitions ($\Theta$).
Note that, aside from these experiments, every other result is based on their default values of $4$ and $16$, respectively.


Second, we evaluate our solution on $k$-way balanced partitioning, the de facto reference for \hgraph partitioning algorithms \cite{AdvancesInHypergraphPartitioning, PartitioningHypergraphsIsHard}.
To do so fairly, we use the standard \mbox{\textbf{cut-net}} quality metric, defined as the total weight of cut \hedges:
\begin{equation}\label{eq:cutnet}
    Cut\text{-}net_G(\rho) = \sum_{e \in E} \omega(e) \cdot \mathbbm{1}(\abs{\{\rho(n) \sothat n \in e\}} > 1) \text{.}
\end{equation}
As for \hgraphs, we use the ISPD98 benchmark \cite{ISPD98}, but given the limited size of its entries, to demonstrate the scalability of massive parallelism, we randomly augment by $16\times$ each \hgraph's nodes and pins, see Tab.~\ref{tab:ispd}.
Here we compare against the multi-threaded CPU partitioner Mt-KaHyPar~\cite{MtKaHyPar} and the GPU partitioner gHyPart~\cite{gHyPart}; we do not include HyperG \cite{HyperG} as no reproducible artifact is available to us.

For our implementation to abide by the $k$-way constraints, we set $\Omega = (1 + \epsilon) \cdot \nicefrac{\abs{N}\,}{k}$ and $\Delta = +\infty$, where $\epsilon$ is a given balance parameter.
Then, to ensure the construction of exactly $k$ initial valid partitions, we halt coarsening upon reaching less than $4096$ coarse nodes (empirically stable for small $k$-s) and rely on Mt-KaHyPar's direct $k$-way configuration \cite{MtKaHyPar} to provide a robust initial partitioning.
Such CPU-side work takes tens of milliseconds and is included in the end-to-end timings.
In our case, the cut-net is not optimized directly, but results from minimizing connectivity.

Experiments were performed on an \texttt{A100-SXM4-40GB} GPU and \texttt{EPYC 7453 @ 2.75GHz} CPU with 256GB of RAM, CUDA version \texttt{12.4}, and GPU driver version \texttt{550.78}.
Every multi-threaded execution was assigned 16 threads.
All timing results are end-to-end and have been averaged over 10 runs with negligible variance.


\subsection{Comparison Results}\label{subsec:comparison_results}







SNN partitioning results are reported in Fig.~\ref{fig:results}.
Our implementation achieves a mean speedup of $380\times$ over hMETIS and $12.5\times$ over the overlap method, remaining within just $1\!\pm\!0.8\times$ of the one-pass method.
Quality of results always improves, with our solution's connectivity on average $0.64\times$ that of hMETIS; an advantage attributable to the larger space of refinement moves explored.
Follow larger improvements of $0.58\times$ and $0.08\times$ over the overlap and one-pass methods.
The number of partitions built presents instead minimal differences, attesting to our effective handling~of~all~constraints.

Results are consistent across all \hgraphs and their topologies.
In particular, sequential execution time exhibits a slight step increase on the less regular \hgraphs (e.g. \texttt{256k-rand}, \texttt{allen-v1}), due to the simultaneous increase in node count and neighbors count per node.
By contrast, our approach's complete parallelism over the \hgraph's size, coupled with the materialization and early deduplication of neighborhoods, lead to a smoother execution time trend.

In Fig.~\ref{fig:results_parallel} we study our solution against Mt-KaHyPar, with the latter allowed to violate the distinct inbound \hedges constraint.
We overall achieve $0.56\times$ Mt-KaHyPar's connectivity, with a speedup ranging from $2.8\times$ to $14.7\times$.
Observably, on three \hgraphs Mt-KaHyPar manages to reach down to $0.79\times$ our connectivity, but at the price of tens of invalid partitions.
On the five largest \hgraphs, notably, the gap in execution time remains between $3$-$9\times$, however, Mt-KaHyPar's connectivity is repeatedly $\sim\!2.0\times$ higher than ours.

We can thus conclude that our approach is capable of consistently achieving SoTA results under the stated constraints, both in terms of quality and time-to-solution.

\subsection{Performance and Scalability Analysis}\label{subsec:performance_and_scalability_analysis}

\begin{figure}[t]
    \centering
    \vspace{-2pt} 
    \includegraphics[width=\columnwidth]{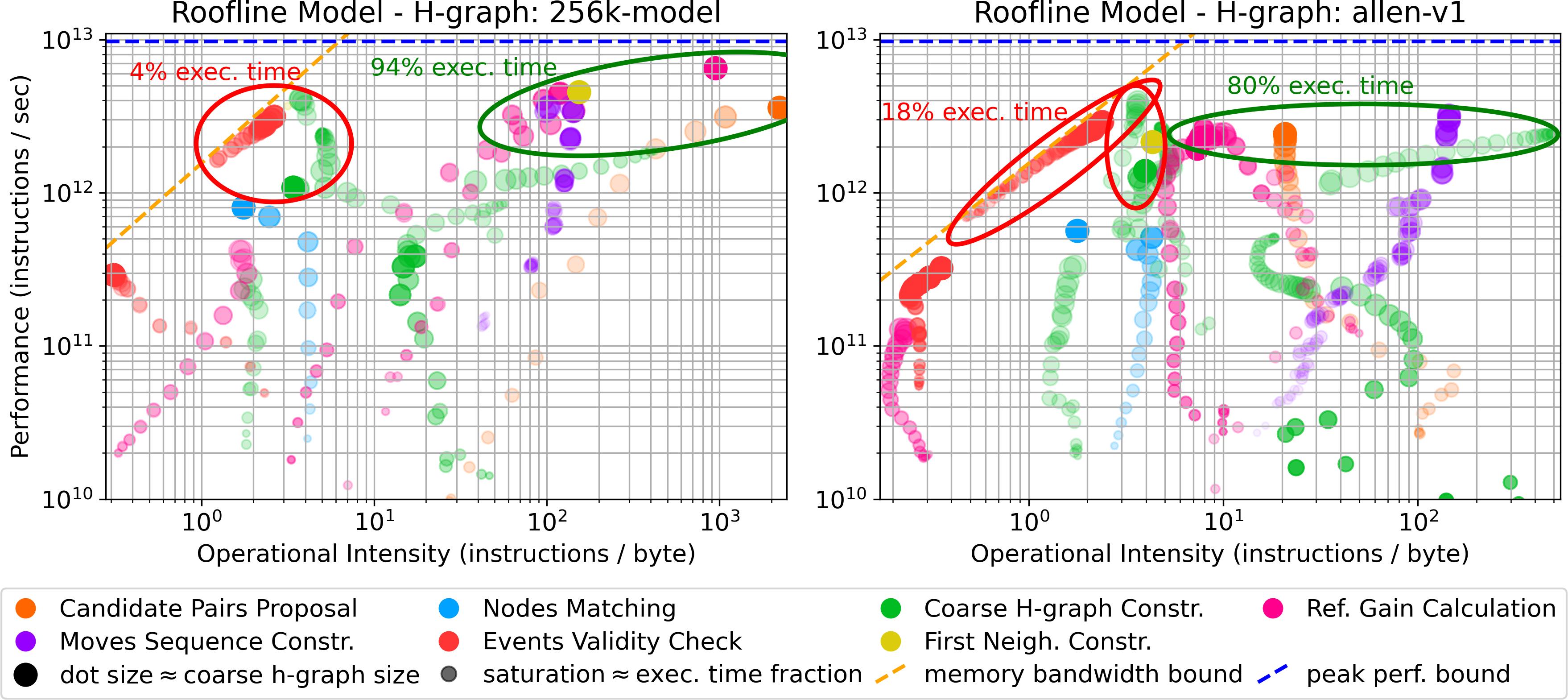}
    \vspace{-14pt} 
    \caption{Integer roofline model for all kernel launches on two SNNs. Each dot is a kernel launch: its size encodes the \hgraph size, its opacity reflects the fraction of execution time within its category.}
    \vspace{-4pt} 
    \label{fig:roofline}
\end{figure}

\begin{figure}[t]
    \centering
    \vspace{-2pt} 
    \includegraphics[width=\columnwidth]{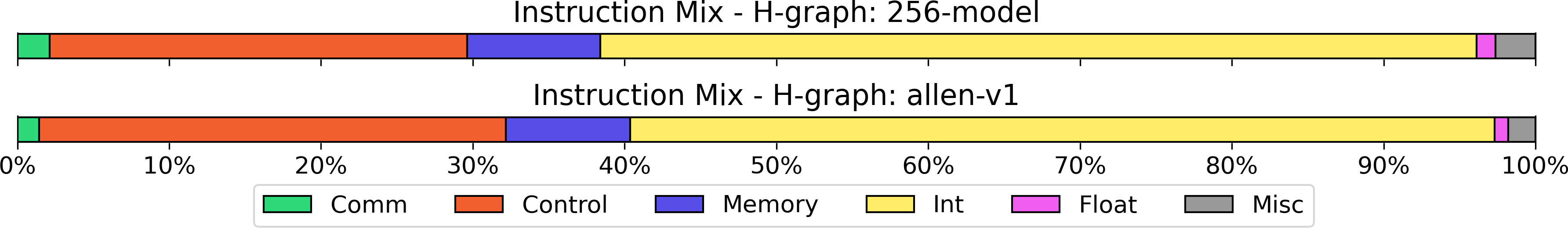}
    \vspace{-14pt} 
    \caption{Instruction mix, aggregated over a complete run.}
    \vspace{-4pt} 
    \label{fig:instruction_mix}
\end{figure}

\begin{figure*}[t]
    \centering
    \vspace{-6pt}
    \includegraphics[width=\textwidth]{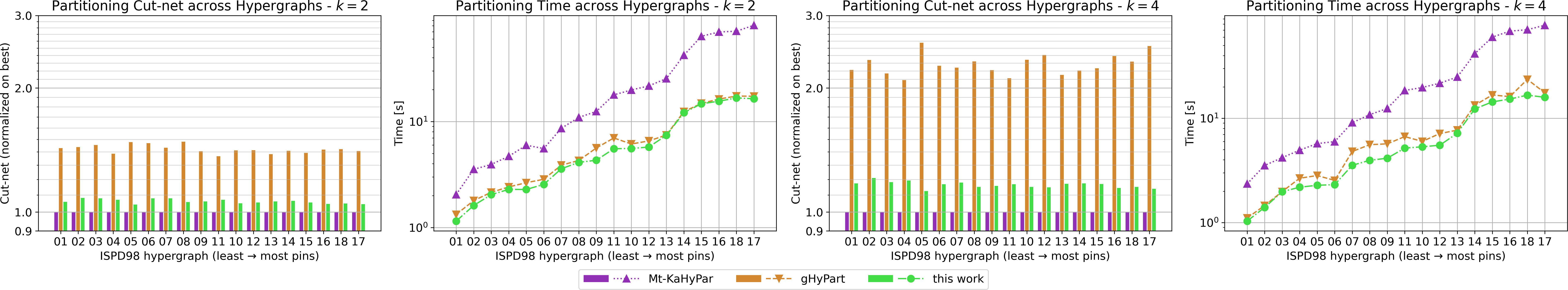}
    \vspace{-13pt} 
    \caption{Balanced $k$-way partitioning results comparison with two parallel methods on the ISPD98 \hgraphs. Constraints: $k = 2,4$, $\epsilon = 0.03$.}
    \vspace{-16pt}
    \label{fig:results_kway}
\end{figure*}

In Tab.~\ref{tab:complexity} we summarize our implementation's complexity analysis.
Such bounds agree with experimental data, as both in Fig.~\ref{fig:results}, and later Fig.~\ref{fig:results_kway}, execution time grows linearly with average \hedge cardinality so long as pins are within tens of millions (e.g. until the 64\kk-model).
This regime is dominated by per-node or per-\hedge work.
Then, as \hgraph size exceeds the GPU's parallelism capacity, additional iterations are serialized within each block and execution time trends linearly with the number of pins.
We note that such a linear growth matches with other GPU implementations \cite{Gkway, HyperG, gHyPart}, hinting at the absence of meaningful overheads from our additional constraints-handling logic.

To quantify the impact of each algorithm step on performance, Fig.~\ref{fig:exec_breakdown} shows how execution time is distributed among them.
As expected, we notice that our two neighborhood iterations, candidates proposal and gain computations, tend to dominate execution on larger inputs \cite{AcceleratedCoarseningProcedure}.
The latter to a lesser extent, thanks to the precomputation of $pins$.
Instead, constraint checks and matching become progressively negligible, as do set operations during coarsening, confirming their ultimate inexpensiveness.
Lastly, the initial construction of neighbors starts off relatively cheap, but as \hedges grow larger, non-unique neighbors grow with $h \cdot d$, making its cost noticeable, but still limited to a one-time overhead.

In our design, host-device memory transfers happen only twice, to load the original \hgraph, and to bring back the finished partitioning.
Initial \hedges and incidence data structures are prepared on the host while importing the \hgraph and copied as is, while neighborhoods and coarse \hgraphs live solely in device memory.
Hence, host-device data movements take up an inconsequential amount of time.

\begin{figure}[t]
    \centering
    \vspace{-2pt} 
    \includegraphics[width=\columnwidth]{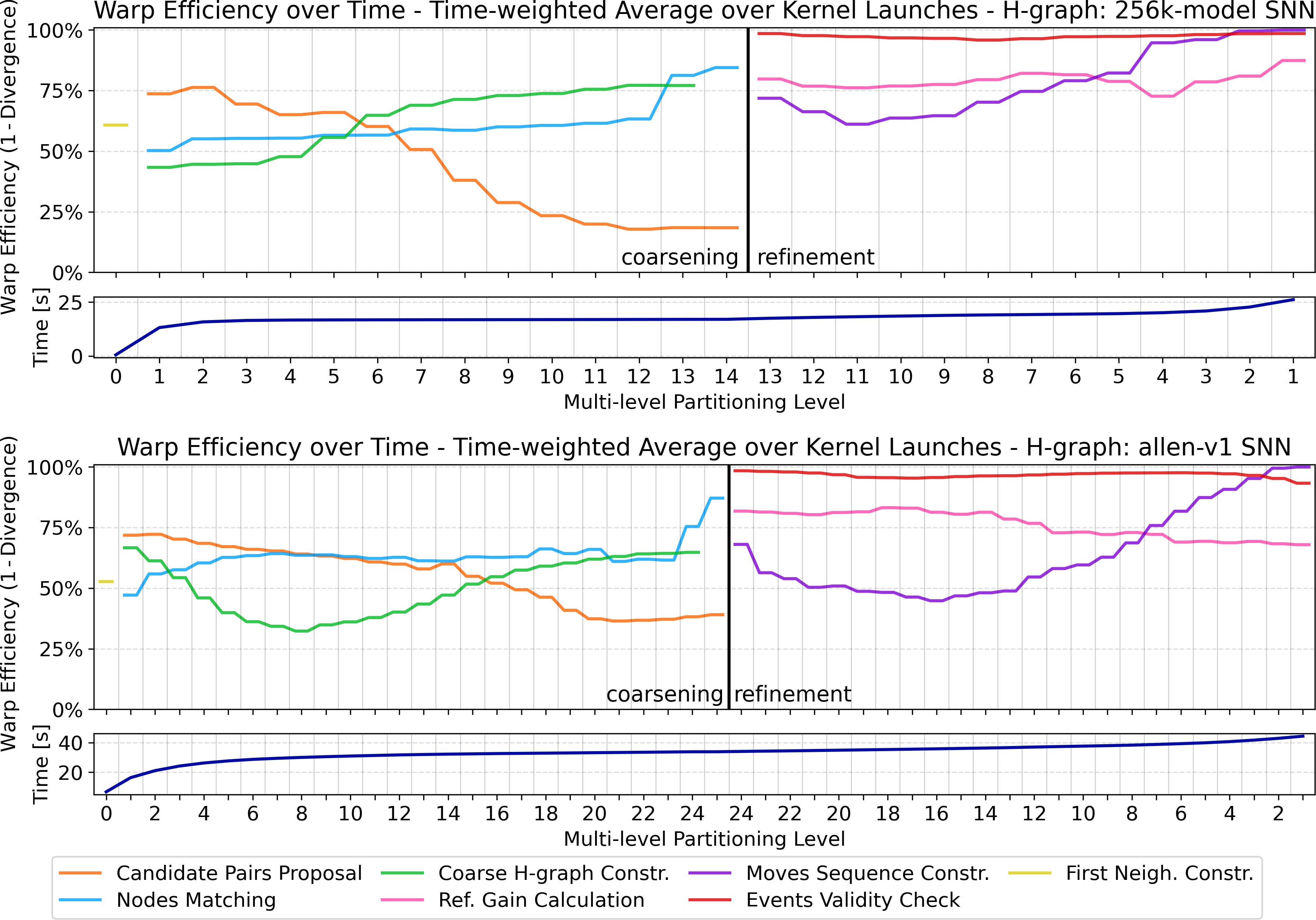}
    \vspace{-14pt} 
    \caption{Average warp efficiency over time on two SNNs.}
    \vspace{-4pt} 
    \label{fig:warp_efficiency}
\end{figure}

\begin{figure}[t]
    \centering
    \vspace{-2pt} 
    \includegraphics[width=\columnwidth]{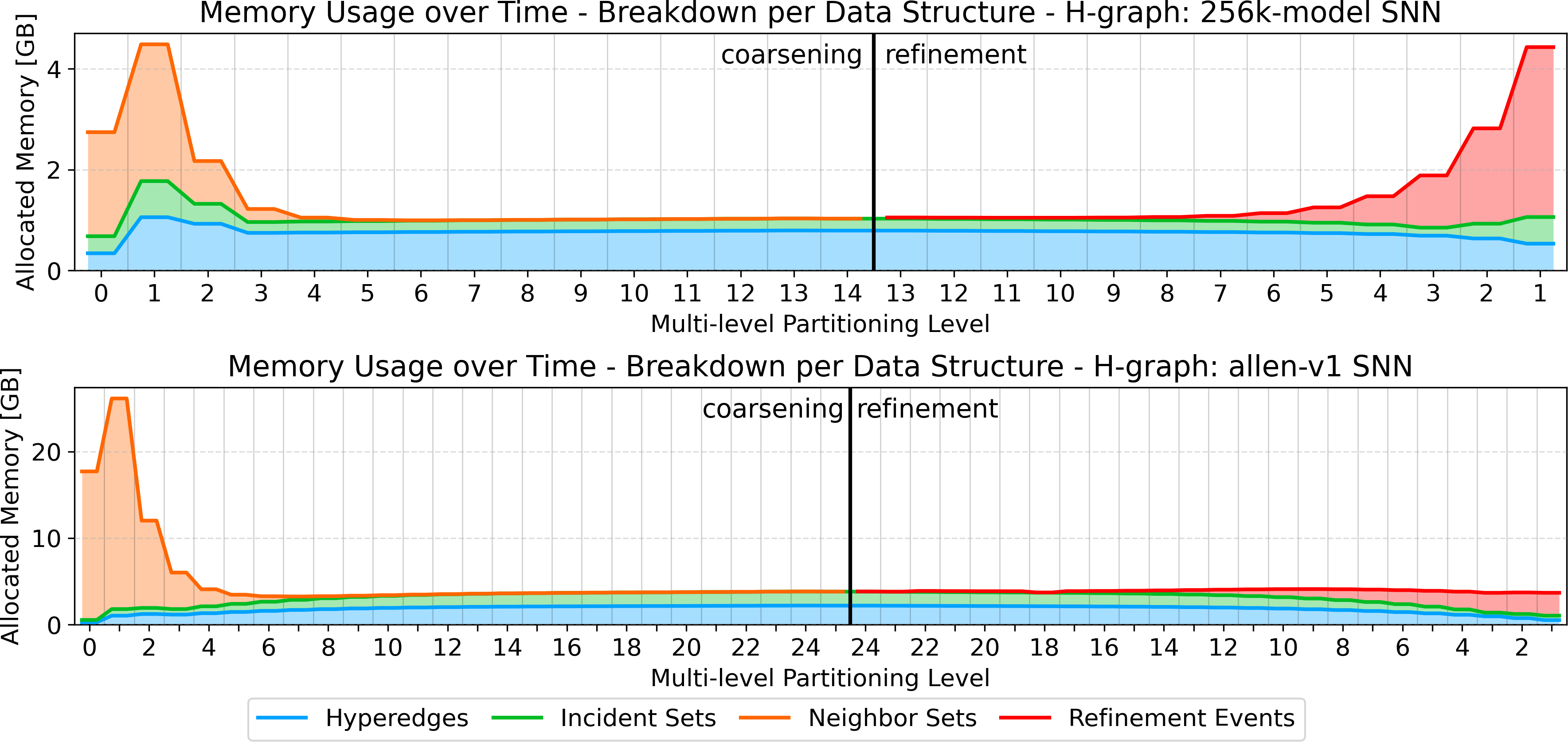}
    \vspace{-14pt} 
    \caption{Memory usage over time on two SNNs.}
    \vspace{-4pt} 
    \label{fig:memory_usage}
\end{figure}

According to the instruction mix in Fig.~\ref{fig:instruction_mix}, the majority of our workload is constituted by integer (\texttt{32bit}) and control instructions, followed by memory accesses.
Floating point operations are a minority, confined to pairing scores and move gains.
This is a natural consequence of \hedges, incidence sets, and neighborhood iterations, whose nested decision-making forms the core of our algorithms.
To assess GPU resource utilization in our design, Fig.~\ref{fig:roofline} presents a roofline model, where each dot corresponds to a kernel launch.
Given our instruction mix, we report results in terms of integer instruction throughput, using DRAM traffic to derive operational intensity.
The intricate nature of \hgraphs and the resulting control flow make it challenging to reach peak performance consistently.
Nonetheless, the vast majority of our global execution time, $\sim\!90\%$, is spent in kernels achieving a substantial fraction of peak throughput, approximately $20\text{–}60\%$.
This is consistent with the fact that only $\sim\!60\%$ of our instructions are integer and directly contribute to the measured throughput.
Low-efficiency kernel launches only amount to a small fraction of runtime and occur on the coarsest levels, where nodes are unavoidably too scarce to reach significant occupancy.


Kernels predominantly operate in a compute-intensive regime, implying that our handling of costly neighborhood iterations exhibits no dominant bottleneck.
Rather, the major factor influencing execution time is just the sheer number of neighbors themselves, whose processing sustains a high level of GPU utilization.
The sole memory-bound steps are the parallel patterns involved in refinement events processing and a few set union operations.
These take up most of the remaining execution time while still reaching non-trivial throughput, typically $>\!10\%$ of peak.

Interestingly, based on the roofline model, the initial neighbors construction (in yellow) should be mostly compute-bound by its large number of iterations, as seen for the \texttt{256k-model}.
Yet, it approaches a memory-bound regime when the topology is more irregular, as observed on the \texttt{allen-v1}.
This can be attributed to a growing number of distinct neighbors and a decrease in duplicates, leading to more updates hitting the global memory hash set.

Warp efficiency ($1-\text{divergence}$), the average fraction of threads participating in the same instruction, is an indicator of proper parallelism exploitation.
We report it in Fig.~\ref{fig:warp_efficiency}.
Most of our kernels use a warp-centric mapping to handle sequential work and enable warp collectives in shared memory (e.g. histogram).
On the initial, large \hgraph, this sustains high intra-warp utilization, $\sim\!70\%$ efficiency for the bulk of runtime.
On inner coarsening levels, the per-node work and neighbor lists shrink, and SIMD efficiency drops below $50\%$; however, these phases contribute to less than $5\%$ of total runtime, as instances are small and finish quickly.
Refinement shows instead a more uniform, high-efficiency trend due to partitions being a simpler decision-target compared to coarsening's neighbors.
Overall, our average efficiency over time always exceeds $70\%$, and optimizing for peak efficiency on small late-stage hypergraphs would have a low impact on the end-to-end time.
These observations further explain and corroborate our previous conclusions from~the~roofline~mode.

Concerning memory utilization, Fig~\ref{fig:memory_usage} presents the amount of memory occupied by different data structures throughout levels.
The baseline memory occupation throughout execution is that of \hedges and inbound sets, with an amount linear in $\abs{N} \cdot d$.
Then, as expected, neighborhoods initially occupy the majority of memory, proportionally to $\abs{N} \cdot d \cdot h$, but that quickly fades after three to four coarsening levels, as they are deduplicated.
In later refinement levels, events too allocate a significant amount of memory, in part due to the support arrays required for their manipulation.
Ultimately, the sole limiting factor to our implementation running under limited VRAM is neighborhoods, one that is easily circumvented by running coarsening kernels in batches over nodes and keeping only a few nodes' neighborhoods in device memory at once.

We report our scalability across GPUs in Fig.~\ref{fig:gpu_scaling}.
There, we see how execution time grows with a similar linear trend regardless of the hardware.
Additionally, the speedup between the \texttt{A100} and \texttt{GH200} increases from $1.26\times$ on the three smallest \hgraphs up to $1.85\times$ on larger ones.
Small \hgraphs underutilize both GPUs, showing only some benefit from architectural improvements.
At scale, however, the speedup is higher than the streaming multiprocessors increase $\nicefrac{144}{108} \approx 1.33\times$ between the two GPUs.
This suggests that, not only is the extra parallelism fully exploited, but also other hardware advantages benefit performance, namely memory bandwidth, cache hierarchy, and scheduling efficiency.
Overall, these results indicate robust scalability with problem size and consistent performance gains across GPU generations.

\slantedispdtable

\begin{figure}[t]
    \centering
    \vspace{-2pt} 
    \includegraphics[width=\columnwidth]{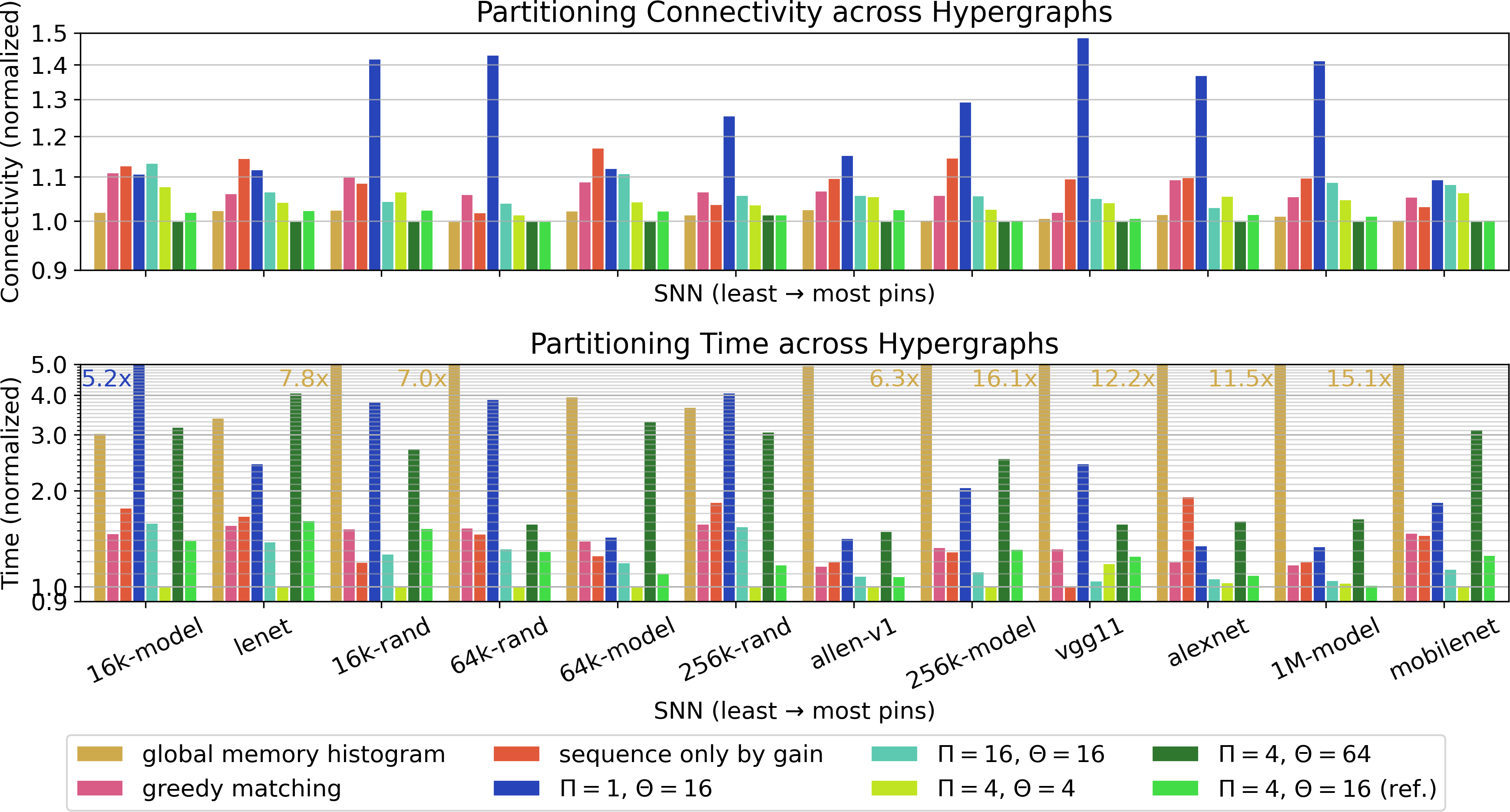}
    \vspace{-14pt} 
    \caption{Ablation study results across twelve SNNs hypergraphs.}
    \vspace{-4pt} 
    \label{fig:results_ablation}
\end{figure}

\subsection{Ablation Studies}\label{subsec:ablation_studies}

Our ablation studies are reported in Fig.~\ref{fig:results_ablation}.
In the first case, we disabled our neighbors materialization, causing the spillage of histograms and deduplication to global memory.
With candidates proposal being our costliest kernel, the impact of such a change more than tripled execution time, reinforcing our choice for materialization.
Secondly, we swapped out our optimal matching for the greedy heuristic in \cite{AxonCUDA-IPDPS}, which always forces roots into a pair and propagates decisions backwards from them.
We thus see that an optimal matching yields an average $0.96\times$ lower connectivity through better coarsening and initial solution, with also a minor speedup thanks to more pairs being constructed, leading to fewer coarsening levels.
At last, we removed our refinement moves chaining step, merely constructing the sequence by gain \cite{HyperG}.
This reduced the effectiveness of refinement under tight constraints and strains events generation, giving both a $1.1\times$ higher connectivity and $1.16\times$ the execution time.
Hence, our decision in favor of chain-based ordering.
Throughout these experiments we kept $\Pi = 4$ and $\Theta = 16$.

Continuing, in terms of our parameters, increasing the number of candidates $\Pi$ from $1$ to $4$ (under $\Theta = 16$) reduces required coarsening levels by $2$-$5$, lowering mean time by $0.5\times$ and connectivity by $0.8\times$.
In particular, the average fraction of matched nodes per level grows from $76\%$ to $94\%$ while initial partitions count drops to $0.84\times$, suggesting that $\Pi = 1$ is too conservative to coarsen effectively.
Going further, to $\Pi = 16$, we see the opposite occur, nearly all nodes are matched per level, time improves by a marginal $0.98\times$, but the exceedingly aggressive coarsening hurts connectivity by $1.05\times$.
All other alternatives for $\Pi \in \{1, \dots 16\}$ were also examined, but $4$ offered the best quality-speed balance.

A higher number of refinement repetitions per level $\Theta$, going from $4$ to $16$ (under $\Pi = 4$), lowers connectivity by an average $0.94\times$, at the cost of $1.21\times$ the execution time.
Reaching for $\Theta = 64$, overall connectivity drops by another $0.98\times$ for a $1.96\times$ longer execution.
Once more, after sweeping every $\Theta \in \{1, \dots 64\}$, we determined that $16$ yielded the highest return on the spent time.
Moreover, under these conditions, the amount of time dedicated to refinement almost matches that of coarsening.

Different configurations for noise threshold and chaining grade parameters (e.g. $\alpha$, $\beta$) were also evaluated, but with negligible effects.

We also considered the use of CUDA Graphs to reduce kernel launch overhead.
However, due to the data-dependent control flow and constantly changing memory layouts across coarsening levels, graph replay provided limited benefit ($1.02\times$ speedup), and was therefore not adopted.

\subsection{Validation on $k$-way Partitioning}\label{subsec:validation_on_k_way_partitioning}

In Fig.~\ref{fig:results_kway} we report the comparison results for $k$-way balanced partitioning.
Starting from $k=2$, $\epsilon = 0.03$.
In terms of speedup over Mt-KaHyPar, our implementation is $1.8\times$ faster on smaller \hgraphs, and progressively grows to a consistent $5\times$ on the four larger ones.
We are instead always $1.0$-$1.3\times$ faster than gHyPart.
Partitioning quality is dominated by Mt-KaHyPar, but our method lags a solid $1.05\times$ behind on average cut-net.
This is to be expected, given the variety of heuristics available on CPU that are incompatible with GPU parallelism, most notably the use of localized backtracking.
Nonetheless, reaching such close results confirms the effectiveness of our algorithms.
On the other hand, our approach consistently improves on gHyPart, with an average $0.75\times$ cut-net.
On $k=4$, same $\epsilon$, we observe a similar pattern in execution time, while cut-net gaps widen slightly to a mean $1.16\times$ between Mt-KaHyPar and our method, and $0.51\times$ between~gHyPart~and~us.


From the above, our solution also proves competitive on $k$-way partitioning.
It achieves near-optimal results in spite of the cut-net not being its primary optimization metric and the burden of extra constraint checks.
At the same time, execution time remains closely aligned with that of an existing GPU partitioner, while substantially improving on its cut-net quality.
These observations testify to the soundness of our parallel implementation of the multi-level scheme.

\section{Conclusion}\label{sec:conclusion}


In this article, we presented the first GPU-centric design and implementation of a deterministic multi-level hypergraph partitioner that supports both partition size and distinct inbound \hedge constraints.
In light of our experiments, it currently achieves among the best results in the field, scaling to hypergraphs with billions of pins with a time-to-solution of tens of seconds.
In fact, execution time grows linearly with the number of pins, while maintaining strong cross-device scalability and efficient GPU resource utilization.
The additional complexity introduced by our particular constraints was subdued with minimal overhead.
Moreover, all major algorithms we proposed in this work have been shown to contribute to the final quality of results; their GPU-friendly design being reflected by the efficient use of hierarchical parallelism.
At last, we demonstrated the generalizability of our solution to $k$-way partitioning, with a compelling balance of speed and results worth.
Hence, advancing support for different constraint sets and optimization metrics constitutes a valuable line of future research.
As of this publication, our implementation is available as open-source \cite{AxonCUDARepo}.

\section*{Acknowledgements}

The authors thank Prof. Sebastiano Schifano and Prof. Cristian Zambelli (Università degli Studi di Ferrara) for providing access to their computing infrastructure, which was used to conduct part of the experiments presented in this work.

This work has been partially supported by the Spoke 1 on \emph{Future HPC} of the Italian Research Center on High-Performance Computing, Big Data and Quantum Computing (ICSC) funded by MUR Mission 4 - Next Generation EU.

\bibliographystyle{IEEEtran}
\bibliography{biblio}

@ARTICLE{ChipletsCombinatorics,
    author={Li, Fuping and Wang, Ying and Lu, Meixuan and Zhu, Yutong and Wang, Haoran and Zhao, Zhun and Huang, Junpei and Wei, Xiaotong and Liang, Xihao and Wang, Yujie and Xu, Haobo and Li, Huawei and Li, Xiaowei and Liu, Qi and Liu, Ming and Sun, Ninghui and Han, Yinhe},
    journal={Integrated Circuits and Systems}, 
    title={The Decomposition and Combination Paradigms of Chiplet-Based Integrated Chips}, 
    year={2024},
    volume={1},
    number={1},
    pages={18-30},
    doi={10.23919/ICS.2024.3451428}
}

@inproceedings{MappingVeryLargeSNNtoNHW,
    author = {Jin, Ouwen and Xing, Qinghui and Li, Ying and Deng, Shuiguang and He, Shuibing and Pan, Gang},
    title = {Mapping Very Large Scale Spiking Neuron Network to Neuromorphic Hardware},
    year = {2023},
    isbn = {9781450399180},
    publisher = {Association for Computing Machinery},
    address = {New York, NY, USA},
    url = {https://doi.org/10.1145/3582016.3582038},
    doi = {10.1145/3582016.3582038},
    booktitle = {Proceedings of the 28th ACM International Conference on Architectural Support for Programming Languages and Operating Systems, Volume 3},
    pages = {419–432},
    numpages = {14},
    location = {Vancouver, BC, Canada},
    series = {ASPLOS 2023}
}

@INPROCEEDINGS{FMpartitioning,
    author={Fiduccia, C.M. and Mattheyses, R.M.},
    booktitle={19th Design Automation Conference}, 
    title={A Linear-Time Heuristic for Improving Network Partitions}, 
    year={1982},
    volume={},
    number={},
    pages={175-181},
    doi={10.1109/DAC.1982.1585498}
}

@article{KaHyPar,
    author = {Schlag, Sebastian and Heuer, Tobias and Gottesb\"{u}ren, Lars and Akhremtsev, Yaroslav and Schulz, Christian and Sanders, Peter},
    title = {High-Quality Hypergraph Partitioning},
    year = {2023},
    issue_date = {December 2022},
    publisher = {Association for Computing Machinery},
    address = {New York, NY, USA},
    volume = {27},
    issn = {1084-6654},
    url = {https://doi.org/10.1145/3529090},
    doi = {10.1145/3529090},
    journal = {ACM J. Exp. Algorithmics},
    month = feb,
    articleno = {1.9},
    numpages = {39}
}

@misc{KaHyPar_origins,
    title={k-way Hypergraph Partitioning via n-Level Recursive Bisection}, 
    author={Sebastian Schlag and Vitali Henne and Tobias Heuer and Henning Meyerhenke and Peter Sanders and Christian Schulz},
    year={2015},
    eprint={1511.03137},
    archivePrefix={arXiv},
    primaryClass={cs.DS},
    url={https://arxiv.org/abs/1511.03137}, 
}

@inproceedings{hMETIS_k_way,
    author = {Karypis, George and Kumar, Vipin},
    title = {Multilevel k-way hypergraph partitioning},
    year = {1999},
    isbn = {1581131097},
    publisher = {Association for Computing Machinery},
    address = {New York, NY, USA},
    url = {https://doi.org/10.1145/309847.309954},
    doi = {10.1145/309847.309954},
    booktitle = {Proceedings of the 36th Annual ACM/IEEE Design Automation Conference},
    pages = {343–348},
    numpages = {6},
    location = {New Orleans, Louisiana, USA},
    series = {DAC '99}
}

@article{hMETIS_vlsi,
    title = "Multilevel hypergraph partitioning: Applications in VLSI domain",
    author = "George Karypis and Rajat Aggarwal and Vipin Kumar and Shashi Shekhar",
    year = "1999",
    doi = "10.1109/92.748202",
    language = "English (US)",
    volume = "7",
    pages = "69--79",
    journal = "IEEE Transactions on Very Large Scale Integration (VLSI) Systems",
    issn = "1063-8210",
    publisher = "Institute of Electrical and Electronics Engineers Inc.",
    number = "1",
}

@ARTICLE{PaToH,
    author={Catalyurek, U.V. and Aykanat, C.},
    journal={IEEE Transactions on Parallel and Distributed Systems}, 
    title={Hypergraph-partitioning-based decomposition for parallel sparse-matrix vector multiplication}, 
    year={1999},
    volume={10},
    number={7},
    pages={673-693},
    doi={10.1109/71.780863}
}

@INPROCEEDINGS{Zoltan,
    author={Devine, K.D. and Boman, E.G. and Heaphy, R.T. and Bisseling, R.H. and Catalyurek, U.V.},
    booktitle={Proceedings 20th IEEE International Parallel \& Distributed Processing Symposium}, 
    title={Parallel hypergraph partitioning for scientific computing}, 
    year={2006},
    volume={},
    number={},
    doi={10.1109/IPDPS.2006.1639359}
}

@inproceedings{BiPart,
    author = {Maleki, Sepideh and Agarwal, Udit and Burtscher, Martin and Pingali, Keshav},
    title = {BiPart: a parallel and deterministic hypergraph partitioner},
    year = {2021},
    isbn = {9781450382946},
    publisher = {Association for Computing Machinery},
    address = {New York, NY, USA},
    url = {https://doi.org/10.1145/3437801.3441611},
    doi = {10.1145/3437801.3441611},
    booktitle = {Proceedings of the 26th ACM SIGPLAN Symposium on Principles and Practice of Parallel Programming},
    pages = {161–174},
    numpages = {14},
    location = {Virtual Event, Republic of Korea},
    series = {PPoPP '21}
}

@inproceedings{PartitioningHypergraphsIsHard,
    author = {Papp, P\'{a}l Andr\'{a}s and Anegg, Georg and Yzelman, Albert-Jan N.},
    title = {Partitioning Hypergraphs is Hard: Models, Inapproximability, and Applications},
    year = {2023},
    isbn = {9781450395458},
    publisher = {Association for Computing Machinery},
    address = {New York, NY, USA},
    url = {https://doi.org/10.1145/3558481.3591087},
    doi = {10.1145/3558481.3591087},
    booktitle = {Proceedings of the 35th ACM Symposium on Parallelism in Algorithms and Architectures},
    pages = {415–425},
    numpages = {11},
    location = {Orlando, FL, USA},
    series = {SPAA '23}
}

@article{AdvancesInHypergraphPartitioning,
    author = {\c{C}ataly\"{u}rek, \"{U}mit and Devine, Karen and Faraj, Marcelo and Gottesb\"{u}ren, Lars and Heuer, Tobias and Meyerhenke, Henning and Sanders, Peter and Schlag, Sebastian and Schulz, Christian and Seemaier, Daniel and Wagner, Dorothea},
    title = {More Recent Advances in (Hyper)Graph Partitioning},
    year = {2023},
    issue_date = {December 2023},
    publisher = {Association for Computing Machinery},
    address = {New York, NY, USA},
    volume = {55},
    number = {12},
    issn = {0360-0300},
    url = {https://doi.org/10.1145/3571808},
    doi = {10.1145/3571808},
    journal = {ACM Comput. Surv.},
    month = mar,
    articleno = {253},
    numpages = {38}
}

@article{gHyPart,
    author = {Wu, Zhenlin and Zhao, Haosong and Liu, Hongyuan and Wen, Wujie and Li, Jiajia},
    title = {gHyPart: GPU-friendly End-to-End Hypergraph Partitioner},
    year = {2025},
    issue_date = {March 2025},
    publisher = {Association for Computing Machinery},
    address = {New York, NY, USA},
    volume = {22},
    number = {1},
    issn = {1544-3566},
    url = {https://doi.org/10.1145/3711925},
    doi = {10.1145/3711925},
    journal = {ACM Trans. Archit. Code Optim.},
    month = mar,
    articleno = {38},
    numpages = {25}
}

@article{MtKaHyPar,
    author = {Gottesb\"{u}ren, Lars and Heuer, Tobias and Maas, Nikolai and Sanders, Peter and Schlag, Sebastian},
    title = {Scalable High-Quality Hypergraph Partitioning},
    year = {2024},
    issue_date = {January 2024},
    publisher = {Association for Computing Machinery},
    address = {New York, NY, USA},
    volume = {20},
    number = {1},
    issn = {1549-6325},
    url = {https://doi.org/10.1145/3626527},
    doi = {10.1145/3626527},
    journal = {ACM Trans. Algorithms},
    month = jan,
    articleno = {9},
    numpages = {54}
}

@misc{AxonCUDA-IPDPS,
    title={Incidence Constraints in Hypergraph Partitioning on GPU}, 
    author={Marco Ronzani and Cristina Silvano},
    year={2026},
    eprint={2604.14411},
    archivePrefix={arXiv},
    primaryClass={cs.DC},
    url={https://arxiv.org/abs/2604.14411},
    note={Accepted at AsHES Workshop @ IPDPS 2026}
}

@misc{AxonFlow,
      title={A Case for Hypergraphs to Model and Map SNNs on Neuromorphic Hardware}, 
      author={Marco Ronzani and Cristina Silvano},
      year={2026},
      eprint={2601.16118},
      archivePrefix={arXiv},
      primaryClass={cs.AR},
      url={https://arxiv.org/abs/2601.16118}, 
}

@Article{AllenV1,
    author={Billeh, Yazan N. and others},
    title={Systematic Integration of Structural and Functional Data into Multi-scale Models of Mouse Primary Visual Cortex},
    journal={Neuron},
    year={2020},
    month={May},
    day={06},
    publisher={Elsevier},
    volume={106},
    number={3},
    pages={388-403.e18},
    issn={0896-6273},
    doi={10.1016/j.neuron.2020.01.040},
    url={https://doi.org/10.1016/j.neuron.2020.01.040}
}

@ARTICLE{LeNet,
    author = {Lecun, Y. and Bottou, L. and Bengio, Y. and Haffner, P.},
    journal = {Proceedings of the IEEE}, 
    title = {Gradient-based learning applied to document recognition}, 
    year = {1998},
    volume = {86},
    number = {11},
    pages = {2278-2324},
    doi = {10.1109/5.726791}
}

@inproceedings{AlexNet,
    author = {Krizhevsky, Alex and Sutskever, Ilya and Hinton, Geoffrey E.},
    title = {ImageNet classification with deep convolutional neural networks},
    year = {2012},
    publisher = {Curran Associates Inc.},
    address = {Red Hook, NY, USA},
    booktitle = {Proceedings of the 26th International Conference on Neural Information Processing Systems - Volume 1},
    pages = {1097–1105},
    numpages = {9},
    location = {Lake Tahoe, Nevada},
    series = {NIPS'12}
}

@misc{VGG,
    title={Very Deep Convolutional Networks for Large-Scale Image Recognition}, 
    author={Karen Simonyan and Andrew Zisserman},
    year={2015},
    eprint={1409.1556},
    archivePrefix={arXiv},
    primaryClass={cs.CV},
    url={https://arxiv.org/abs/1409.1556}, 
}

@misc{MobileNet,
    title={MobileNets: Efficient Convolutional Neural Networks for Mobile Vision Applications}, 
    author={Andrew G. Howard and Menglong Zhu and Bo Chen and Dmitry Kalenichenko and Weijun Wang and Tobias Weyand and Marco Andreetto and Hartwig Adam},
    year={2017},
    eprint={1704.04861},
    archivePrefix={arXiv},
    primaryClass={cs.CV},
    url={https://arxiv.org/abs/1704.04861}, 
}

@inproceedings{HyperG,
    author = {Lee, Wan Luan and Lin, Dian-Lun and Chiu, Cheng-Hsiang and Schlichtmann, Ulf and Huang, Tsung-Wei},
    title = {HyperG: Multilevel GPU-Accelerated k-way Hypergraph Partitioner},
    year = {2025},
    isbn = {9798400706356},
    publisher = {Association for Computing Machinery},
    address = {New York, NY, USA},
    url = {https://doi.org/10.1145/3658617.3697551},
    doi = {10.1145/3658617.3697551},
    booktitle = {Proceedings of the 30th Asia and South Pacific Design Automation Conference},
    pages = {1031–1040},
    numpages = {10},
    location = {Tokyo, Japan},
    series = {ASPDAC '25}
}

@inproceedings{Gkway,
    author = {Lee, Wan Luan and Lin, Dian-Lun and Huang, Tsung-Wei and Jiang, Shui and Ho, Tsung-Yi and Lin, Yibo and Yu, Bei},
    title = {G-kway: Multilevel GPU-Accelerated k-way Graph Partitioner},
    year = {2024},
    isbn = {9798400706011},
    publisher = {Association for Computing Machinery},
    address = {New York, NY, USA},
    url = {https://doi.org/10.1145/3649329.3656238},
    doi = {10.1145/3649329.3656238},
    booktitle = {Proceedings of the 61st ACM/IEEE Design Automation Conference},
    articleno = {105},
    numpages = {6},
    location = {San Francisco, CA, USA},
    series = {DAC '24}
}

@INPROCEEDINGS{SIMDefficientGraphsOnGPU,
    author={Khorasani, Farzad and Gupta, Rajiv and Bhuyan, Laxmi N.},
    booktitle={2015 International Conference on Parallel Architecture and Compilation (PACT)}, 
    title={Scalable SIMD-Efficient Graph Processing on GPUs}, 
    year={2015},
    volume={},
    number={},
    pages={39-50},
    doi={10.1109/PACT.2015.15}
}

@INPROCEEDINGS{AcceleratedCoarseningProcedure,
    author={Lin Cheng and Hyunsu Cho and Yoon, Peter},
    booktitle={2015 IEEE High Performance Extreme Computing Conference (HPEC)}, 
    title={An accelerated procedure for hypergraph coarsening on the GPU}, 
    year={2015},
    volume={},
    number={},
    pages={1-7},
    doi={10.1109/HPEC.2015.7322449}
}

@inproceedings{TreesOnGPU,
    author = {Gupta, Chetan and Latypov, Rustam and Maus, Yannic and Pai, Shreyas and S\"{a}rkk\"{a}, Simo and Studen\'{y}, Jan and Suomela, Jukka and Uitto, Jara and Vahidi, Hossein},
    title = {Fast Dynamic Programming in Trees in the MPC Model},
    year = {2023},
    isbn = {9781450395458},
    publisher = {Association for Computing Machinery},
    address = {New York, NY, USA},
    url = {https://doi.org/10.1145/3558481.3591098},
    doi = {10.1145/3558481.3591098},
    booktitle = {Proceedings of the 35th ACM Symposium on Parallelism in Algorithms and Architectures},
    pages = {443–453},
    numpages = {11},
    location = {Orlando, FL, USA},
    series = {SPAA '23}
}

@article{HypergraphForSparseMatMul,
    author = {Ballard, Grey and Druinsky, Alex and Knight, Nicholas and Schwartz, Oded},
    title = {Hypergraph Partitioning for Sparse Matrix-Matrix Multiplication},
    year = {2016},
    issue_date = {December 2016},
    publisher = {Association for Computing Machinery},
    address = {New York, NY, USA},
    volume = {3},
    number = {3},
    issn = {2329-4949},
    url = {https://doi.org/10.1145/3015144},
    doi = {10.1145/3015144},
    journal = {ACM Trans. Parallel Comput.},
    month = dec,
    articleno = {18},
    numpages = {34}
}

@article{ChallengesInDynamicLoadBalancing,
    title = {New challenges in dynamic load balancing},
    journal = {Applied Numerical Mathematics},
    volume = {52},
    number = {2},
    pages = {133-152},
    year = {2005},
    note = {ADAPT '03: Conference on Adaptive Methods for Partial Differential Equations and Large-Scale Computation},
    issn = {0168-9274},
    doi = {https://doi.org/10.1016/j.apnum.2004.08.028},
    url = {https://www.sciencedirect.com/science/article/pii/S0168927404001631},
    author = {Karen D. Devine and Erik G. Boman and Robert T. Heaphy and Bruce A. Hendrickson and James D. Teresco and Jamal Faik and Joseph E. Flaherty and Luis G. Gervasio}
}

@BOOK{ParameterizedAlgorithms,
    title = "Parameterized Algorithms",
    author = "Cygan, Marek and Fomin, Fedor V and Marx, Daniel and Saurabh, Saket and Kowalik, Lukasz and Lokshtanov, Daniel and Pilipczuk, Marcin",
    publisher = "Springer International Publishing",
    edition = 1,
    month = jul,
    year = 2015,
    address = "Cham, Switzerland",
    copyright = "https://www.springernature.com/gp/researchers/text-and-data-mining",
    language = "en"
}

@inproceedings{ISPD98,
    author = {Alpert, Charles J.},
    title = {The ISPD98 circuit benchmark suite},
    year = {1998},
    isbn = {158113021X},
    publisher = {Association for Computing Machinery},
    address = {New York, NY, USA},
    url = {https://doi.org/10.1145/274535.274546},
    doi = {10.1145/274535.274546},
    booktitle = {Proceedings of the 1998 International Symposium on Physical Design},
    pages = {80–85},
    numpages = {6},
    location = {Monterey, California, USA},
    series = {ISPD '98}
}

@misc{AxonCUDARepo,
    author = {Ronzani, Marco},
    title = {open-source artifact},
    year = {2026},
    howpublished = {\url{https://github.com/EMJzero/AxonCUDA}},
    note = {Accessed: 2026-04-01}
}

@dataset{BenchmarkSNNs,
    author = {Ronzani, Marco},
    title = {Spiking Neural Network Hypergraphs with Spike Frequency Data},
    month = mar,
    year = 2026,
    publisher = {Zenodo},
    version = {v0.1.0},
    doi = {10.5281/zenodo.19194881},
    url = {https://doi.org/10.5281/zenodo.19194881},
}

\begin{IEEEbiography}[{
    \includegraphics[width=1in,height=1.25in,clip,keepaspectratio]{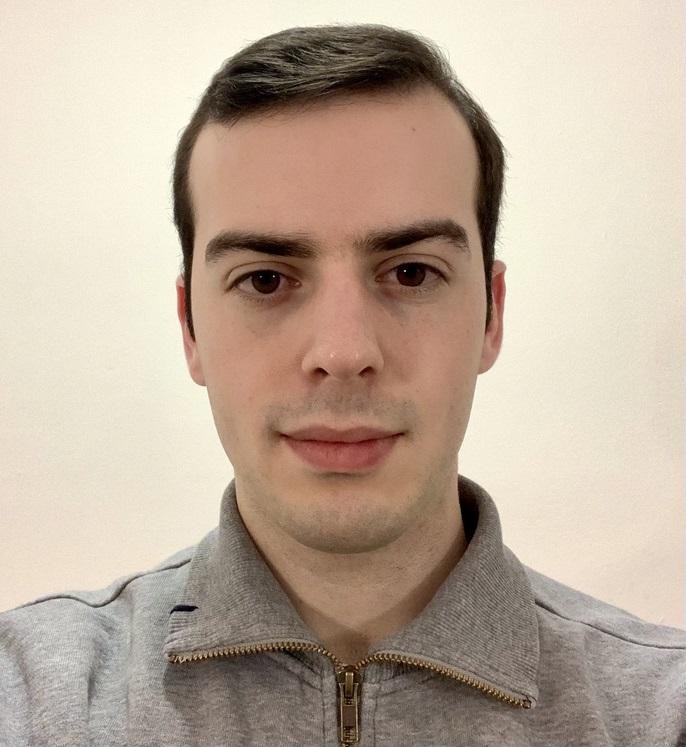}
}]{Marco Ronzani} received the B.S. and M.S. degrees in Computer Science and Engineering from Politecnico di Milano, Italy, in 2022 and 2024, respectively.
He is currently pursuing the Ph.D. degree at the same institution.
His research interests include algorithm engineering for parallel and GPU computing, with a focus on the runtime optimization of hardware accelerators through large-scale combinatorial methods.
His current work explores hypergraph algorithms for neuromorphic computing systems.
\end{IEEEbiography}


\begin{IEEEbiography}[{
    \includegraphics[width=1in,height=1.25in,clip,keepaspectratio]{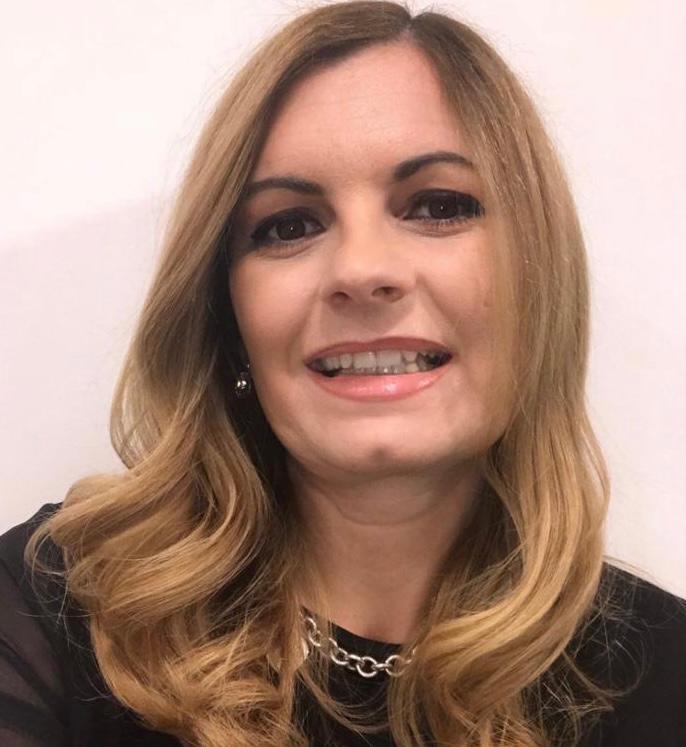}
}]{Cristina Silvano} is a Full Professor of Computer Architecture at Politecnico di Milano, where she is the Chair of the Research Area on Computer Science and Engineering.
In 2022, she was one of the promoters of the new M.Sc. degree in HPC Engineering at Politecnico di Milano. 
Since 2017, she is an IEEE Fellow for contributions to energy-efficient computer architectures.
Her research activities are in the areas of computer architecture and EDA, with emphasis on design space exploration of energy-efficient architectures, low-power design of manycore architectures, accelerators for deep neural networks, and application autotuning for HPC.
She has published more than 200 peer-reviewed papers, six books, and some patents.
She has been Scientific Coordinator of three European research projects (ANTAREX, 2PARMA and MULTICUBE).
She is an active member of the scientific community and she regularly serves in several international program committees.
She is Associate Editor of the ACM Trans. on Computer Architecture and Compiler Optimization and Associate Editor-in-Chief of the Journal on Parallel and Distributed Computing. She serves regularly as independent expert reviewer for the European Commission and for several national science foundations.
In 2022, she was in the National Expert Group on Semiconductor Technologies appointed by the Italian Ministry of University and Research.
\end{IEEEbiography}

\end{document}